\documentclass[sigconf,natbib=true,anonymous=false]{acmart}

\usepackage[utf8]{inputenc}
\usepackage{newfloat}
\usepackage{listings}
\usepackage{subcaption}
\usepackage{caption}
\usepackage{multirow}
\usepackage{hyperref}
\usepackage{xcolor}
\usepackage{pifont}
\usepackage{colortbl}
\usepackage{adjustbox}
\usepackage{array, makecell}
\usepackage{tikz}

\AtBeginDocument{%
  \providecommand\BibTeX{{%
    \normalfont B\kern-0.5em{\scshape i\kern-0.25em b}\kern-0.8em\TeX}}}

\setcopyright{acmcopyright}
\copyrightyear{2023}
\acmYear{2023}
\acmDOI{XXX}

\acmConference[SIGIR '23]{The 46th International ACM SIGIR Conference on Research and Development in Information Retrieval}{July 23--27,2023}{Taipei, TaiWan}
\acmPrice{15.00}
\acmISBN{978-1-4503-XXXX-X/18/06}

\begin{document}

\title{Form-NLU: Dataset for the Form Language Understanding}

\author{Yihao Ding}
\affiliation{%
  \institution{The University of Sydney}
  \city{Sydney}
  \state{NSW}
  \country{Australia}
}

\author{Siqu Long}
\affiliation{%
  \institution{The University of Sydney}
  \city{Sydney}
  \state{NSW}
  \country{Australia}
}

\author{Jiabin Huang}
\affiliation{%
  \institution{The University of Sydney}
  \city{Sydney}
  \state{NSW}
  \country{Australia}
}

\author{Kaixuan Ren}
\affiliation{%
  \institution{The University of Sydney}
  \city{Sydney}
  \state{NSW}
  \country{Australia}
}

\author{Xingxiang Luo}
\affiliation{%
  \institution{The University of Sydney}
  \city{Sydney}
  \state{NSW}
  \country{Australia}
}

\author{Hyunsuk Chung}
\affiliation{%
  \institution{FortifyEdge}
  \city{Sydney}
  \state{NSW}
  \country{Australia}
}

\author{Soyeon Caren Han}

\affiliation{%
  \institution{The University of Sydney}
  \city{Sydney}
  \state{NSW}
  \country{Australia}
}

\renewcommand{\shortauthors}{Ding, et al.}

\begin{abstract}
Compared to general document analysis tasks, form document structure understanding and retrieval are challenging. Form documents are typically made by two types of authors; A form designer, who develops the form structure and keys, and a form user, who fills out form values based on the provided keys. Hence, the form values may not be aligned with the form designer's intention (structure and keys) if a form user gets confused. In this paper, we introduce Form-NLU, the first novel dataset for form structure understanding and its key and value information extraction, interpreting the form designer's intent and the alignment of user-written value on it. It consists of 857 form images, 6k form keys and values, and 4k table keys and values. Our dataset also includes three form types: digital, printed, and handwritten, which cover diverse form appearances and layouts. We propose a robust positional and logical relation-based form key-value information extraction framework. Using this dataset, Form-NLU, we first examine strong object detection models for the form layout understanding, then evaluate the key information extraction task on the dataset, providing fine-grained results for different types of forms and keys. Furthermore, we examine it with the off-the-shelf pdf layout extraction tool and prove its feasibility in real-world cases. 
\end{abstract}






\maketitle
\section{Introduction}
The structural information and its value extraction in a document can be a valuable source for Natural Language Processing (NLP) tasks, especially information extraction and retrieval. Recently, NLP communities and industries, like IBM and Microsoft, have proposed a range of techniques to \textit{understand} positional and/or logical structure of documents\cite{publaynet,docparser,zhang2021entity,docgcn}, and \textit{extract} the essential information\cite{vdoc}. Those researches have mainly focused on Visually Rich Document(VRD)-based tasks, such as academic papers\cite{publaynet, docparser}, receipts\cite{scoie, cord}, and forms \cite{funsd, xfund}, and many benchmark problems has been solved, including layout analysing \cite{publaynet,docbank}, table structure recognition \cite{pubtabnet, scitsr, fintabnet}, document question answering \cite{docvqa, visualmrc}. 

Most VRD-based problems have been successfully solved, however, the form-understanding task is relatively challenging. This is mainly because of two reasons: two types of authors in a form and the combination of diverse visual cues. First, unlike general VRDs, the main aim of forms is very clear, \textit{collecting data from the form users}, and this aim is applied from the medical domain to administrative data collection. According to this aim, we can assume that there are two main authors; A form designer and a form user. A form designer focuses on developing a form structure to collect the required information by defining the clear key point. The developed form would be used as a user interface so a form user can supply the form value based on their understanding.  Unfortunately, not every form is clear and easy to understand. To collect the diverse required information, several form designers tend to make forms with diverse layouts/structures, which have complex logical and positional relationships between semantic entities. The form user can easily get confused with the designer's intention, and this derives a wrong alignment of key-value pairs. The confusion about the form designer's intention and the uncertainty of the form user would raise the difficulty of form document understanding and information extraction. Secondly, due to the involvement of \textit{form developers-to-users} relationship, the form has a high possibility of having a combination of different natures, such as digital, printed, or handwritten. For example, a designer may provide users with electrical paper forms, while users may submit filled forms via various carriers, such as digital, printed or handwritten versions. It also commonly happens that users provide various types of noise (low resolution or uneven scanning or bad handwriting) in the submitted forms. This causes huge difficulties in understanding form document structure and extracting the essential key-value pairs. Regarding form understanding, several datasets have been released in recent years, collected from scanned receipts \cite{cord,scoie}, contracts\cite{kleister}, and cross-domain forms \cite{funsd,xfund} (shown in Table~\ref{tab:datasets}). However, those datasets produce the form developer's intention as relatively simple and general, which does not deal with the confusion about the form designer's intention and the form user's uncertainty. Moreover, most datasets do not cover the various carriers of document versions and their noises. This would worsen understanding of the form structure and extracting the key information. 

In this paper, we introduce a new dataset for form structure understanding and key information extraction. The dataset would enable the interpretation of the form designer's specific intention and the alignment of user-written value on it. Our dataset also includes three form types: digital, printed, and handwritten, which cover diverse form appearances/layouts and deal with their noises. 
In addition to this, we propose a new baseline for form structure understanding and key-value information extraction, which applies robust positional and logical relations. To do this, we cover user-caused form diversities to precisely extract key information from forms, which is an emerging industrial demand currently. Our model covers the hierarchical structure of documents, from words to sentences, sentences to a semantic entities like a paragraph, and entities to a document page. 
Note that exiting transformer-based models mainly focus on token \cite{xu2020layoutlm, xu2021layoutlmv2, huang2022layoutlmv3, xu2021layoutxlm,gu2022xylayoutlm} or entity level \cite{zhang2021entity,tan2019lxmert,visualbert} independently, ignoring the contextual dependency between different level elements.  
In the evaluation, we first examine strong object detection models for the form layout understanding, then test the proposed model for the key information extraction task. Moreover, we also examine our key information extraction dataset and proposed model with the off-the-shelf pdf layout extraction tool and prove the feasibility in real-world cases. 

The main contribution of this research can be summarised as follows: 1) We introduce Form-NLU, a new form structure understanding and key information extraction dataset that covers specific form designers' intentions and makes an alignment with the user's values. 2) We propose a new baseline model that handles positional and logical relations and hierarchical structure in form documents. The proposed model has outperformed other SOTA form key information extraction models on Form-NLU. 3) We apply the proposed dataset and the model with the off-the-shelf pdf layout extraction tool and prove the feasibility in real-world form document cases.
\section{Related Work}
There are general documents understanding benchmarks introduced in \cite{publaynet, docparser, visualmrc}, but we briefly introduce the form document understanding and information extraction, which involves multi-party interaction resulting in more complicated positional and logical relationships, as shown in Table \ref{tab:datasets}.
There are three major points that we would like to discuss and compare with previous benchmarks. First, most benchmarks \cite{funsd, xfund, ephoie} do not cover the various carriers of document versions and their noises. For example, a form designer may provide the form with digital forms, while the form user may submit the filled form with various carriers, such as digital ($\mathcal{D}$), printed ($\mathcal{P}$) or handwritten($\mathcal{H}$) version. This trend commonly affects the quality of form structure understanding and information extraction due to the various types of noise, including resolution or scanning issues. Hence, the successful benchmark should cover real-world cases with various carriers. 
Secondly, in order to conduct the form structure understanding and information extraction, it is crucial to interpret the positional and logical relationships between form components. Most benchmarks enable understanding positional and logical/semantic relations by using bounding boxes (B.Box) and Textual information. However, \citet{kleister} releases two form information extraction datasets, NDA and Charity, without providing bounding box coordinates of semantic entities. 
Finally, the form document has two main authors, form designers and form users. The form designers design the form structure to collect the required information (designer's intent), and the form users try to understand the designer's intention but easily get confused. Hence, handling the designer's intention is crucial to deal with understanding the form structure and extracting key information. Several popular benchmarks \cite{funsd, xfund} produce the form developer's intention as relatively simple and general, which does not deal with the confusion about the form designer's intention and the form user's uncertainty. For example, those datasets cover simple form component types, keys and values. Scanned exam paper \cite{ephoie} datasets 
have a relatively simple layout with horizontal key-value pair structures, which do not include any dynamic layout components, such as tables, paragraphs, or complex key-value pairs. 
\textbf{Form-NLU} is the first visual-linguistics form language understanding dataset for supporting researchers in interpreting specific designer intentions under noises from user's input with various types of form carriers.

\begin{table}[t]
\begin{center}
  \begin{adjustbox}{max width=0.48\textwidth}
   
  \begin{tabular}
  {l|l|lll|ll|c}
  
  \hline
     \multirow{2}{*}{\textbf{Name}}& \multirow{2}{*}{\textbf{Source}} &\multicolumn{3}{c|}{\textbf{Type}}& \multicolumn{2}{c|}{\textbf{Features}} &  \multicolumn{1}{c}{\textbf{Designer}}\\
  \cline{3-7}
         &  &  $\mathcal{D}$ & $\mathcal{P}$ & $\mathcal{H}$ & \textbf{B.Box} &\textbf{Text} & \bf Intention\\
     \hline
     FUNSD~\citep{funsd} &Noise Form\ &$\times$ & $\bigcirc$ & $\bigcirc$ &  $\bigcirc$ & $\bigcirc$ & General\\
     XFUND~\cite{xfund} &Synthetic Form\ &$\bigcirc$ & $\times$ & $\bigcirc$ &    $\bigcirc$ & $\bigcirc$ & General\\
     EPHOIE~\cite{ephoie}& Exam Paper& $\times$ & $\times$ &$\bigcirc$ &    $\bigcirc$ & $\bigcirc$ & General\\
     Charity~\cite{kleister} & Annual Report  & $\bigcirc$ & $\bigcirc$ &$\bigcirc$ &   $\times$ & $\bigcirc$ & Specific\\
     NDA~\cite{kleister} & Agreements  & $\bigcirc$ & $\times$ &$\times$ &    $\times$ & $\bigcirc$ & Specific\\
     \hline
     Form-NLU \textbf{(Ours)} &Financial Form\ &$\bigcirc$ & $\bigcirc$ & $\bigcirc$ &   $\bigcirc$ & $\bigcirc$ & Specific\\
     \hline
 \end{tabular}
 \end{adjustbox}
    \caption{Summary of Form Understanding Datasets.}
 \label{tab:datasets}
 \vspace*{-2em}
 \end{center}
 \end{table}

\section{Dataset}
\subsection{Data Collection}
\begin{figure}[!t]
    \hspace*{0cm}
     \centering
     \hspace*{0em}
     \begin{subfigure}[b]{0.15\textwidth}
         \centering
         \includegraphics[height=3.3cm]{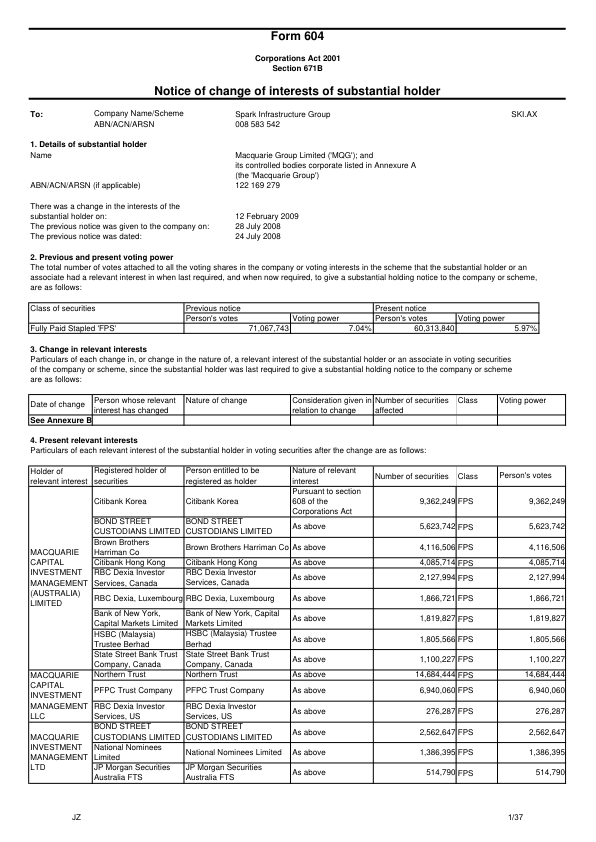}
         \caption{Digital ($\mathcal{D}$)}
         \label{fig:digital_sample}
     \end{subfigure}
     \hspace*{0em}
     \begin{subfigure}[b]{0.15\textwidth}
         \centering
         \includegraphics[height=3.3cm]{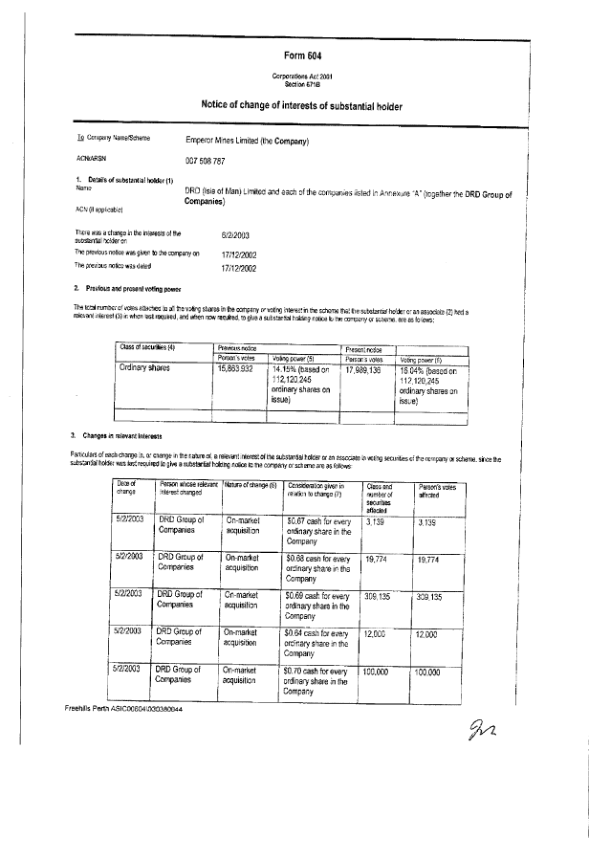}
         \caption{Printed ($\mathcal{P}$)}
         \label{fig:printed_sample}
     \end{subfigure}
     \hspace*{0em}
     \begin{subfigure}[b]{0.15\textwidth}
         \centering
         \includegraphics[height=3.3cm]{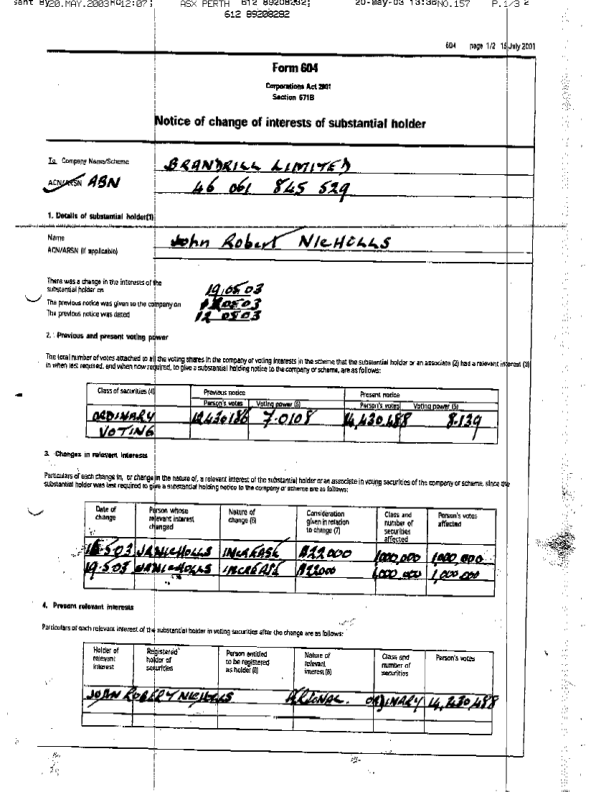}
         \caption{Handwritten ($\mathcal{H}$)}
         \label{fig:handwritten_sample}
     \end{subfigure}
        \caption{Digital, Printed and Handwritten Form Samples.}
        \label{fig:form_samples}
\end{figure}

Our \textbf{Form-NLU} is a subset of the publicly available financial form data source for Form 604 (notice of change of interests of the substantial holder)\footnote{\url{https://asic.gov.au/regulatory-resources/forms/forms-folder/604-notice-of-change-of-interests-of-substantial-holder/}} collected by SIRCA. It provides the text records of each substantial shareholder notice form submitted to the Australian Stock Exchange (ASX)\footnote{\url{https://www2.asx.com.au/}} from 2003 to 2015. To better comprehend the form designer's intentions, we included the twelve most essential form fields~\cite{chang2016informativeness}. Each form field expects a \textbf{Value} from the user, which should contain the specific information being asked by the corresponding \textbf{Key} that expresses the form designer's intention. For example, the key of \textit{"company name"} anticipates a string value that gives the name of a company, whereas \textit{"voting percentage"} asks for a  number that tells the portion of the voting. In order to address the aforementioned developer-to-user relationships in real-world scenarios, we also covered three variants of format depending on how the forms were filled and submitted by different users: (1) Digital ($\mathcal{D}$), (2) Printed ($\mathcal{P})$ and (3) Handwritten ($\mathcal{H}$). The Digital forms are filled in digitally and directly submitted as PDF files. The Printed and Handwritten forms are filled in digitally and by hand respectively and then scanned first before being saved as PDF files for submission. A sample of each type of the form is provided in Figure~\ref{fig:form_samples}, which demonstrates the potential diversity in terms of the form format, e.g., font styles and rotations.  

\begin{figure}[t]
     \centering
         \includegraphics[width=0.46\textwidth]{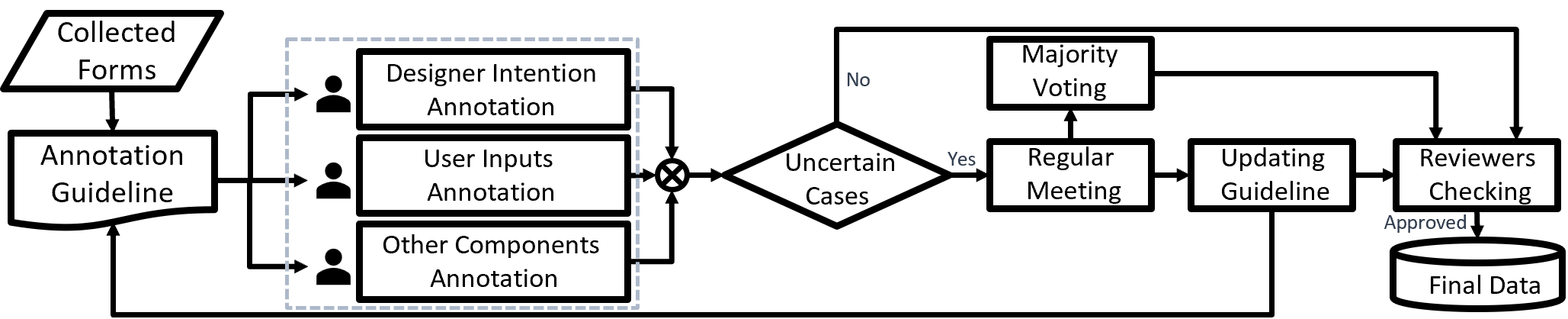}
         \caption{Overall Annotation Workflow.}
         \label{fig:annotation_procedure}
          \vspace*{-1em}
\end{figure}

\subsection{Annotation Guideline}
Based on our task goals (see Section \ref{sec:task_overview}), we first developed the annotation schema and guidelines. For Layout Analysis (as Task A\footnote{See Section \ref{sec:taskAdef} Task A - Form Layout Analysing}), we created seven distinct form layout component categories (\textit{Title}, \textit{Section}, \textit{Form Key}, \textit{Form Value}, \textit{Table Key}, \textit{Table Value} and \textit{Others}) that encompass all possible components from the collected forms. Each form layout component represents a semantic segment of the text area. The \textit{Title} and \textit{Section} correspond to the title of the form document and the sections, respectively. The \textit{Form/Table Key} and \textit{Form/Table Value} refer to the intended questions of the form designer (e.g. Company Name) and the corresponding user input fields. The difference between the \textit{Key/Value} of form and table is that, by our definition, the \textit{Form Key-Value} pairs are horizontally aligned, whereas the \textit{Table Key-Value} pairs are vertically aligned. Any components that do not fall into these four types are defined as \textit{Others}, such as a customized paragraph and cells added and modified by some companies on their own purposes. 

While the Layout Anlysis (as Task A) focuses on understanding the overall form structure via those general semantic components, the Key Information Extraction (as Task B\footnote{See Section \ref{sec:taskBdef} Task B - Key Information Extraction}) aims to comprehend different intentions of the form designers. Thus, in Task B, the \textit{Key} and \textit{Value} components from Task A are further distinguished based on the twelve Key intentions, which include: (1) Company Name (\textit{com\_nm}), (2) Company ID (\textit{com\_id}), (3) Holder Name (\textit{hold\_nm}), (4) Holder ID (\textit{hold\_id}), (5) Change Date (\textit{chg\_date}), (6) Previous Notice Given Date (\textit{gvn\_date}), (7) Previous Notice Date (\textit{ntc\_date}), (8) Class of Securities (\textit{class}), (9) Previous Share (\textit{pre\_shr}), (10) Previous Voting Percentage (\textit{pre\_pct}), (11) Share (\textit{new\_shr}), (12) New Voting Percentage (\textit{new\_pct}). \textit{Key} (1)-(2) and \textit{Key} (3)-(4) ask for the identity of the listed company and the substantial holder, respectively, whereas \textit{Key} (5)-(12) query the value of the shares being changed. We simply appended the identifier \textit{Key} and \textit{Value} separately to the twelve(12) \textit{Key} names, and created twenty-four(24) labels for the key-value pairs accordingly, e.g., Each key-value pair would produce 2 labels; for example, \textit{"Company Name Key"} and \textit{"Company Name Value"}. Based on these different key intentions, we inspected the value patterns to explore the potential noise caused by the diversified user understanding and uncertainty. The statistical analysis can be found in Section \ref{sec:dataanalysis} (Figure~\ref{fig:key value pattern}).

\subsection{Annotation Procedure}
We recruited three human annotators and two human reviewers\footnote{The three annotators are with a background in Computer Science or Financial at the University of Sydney while the two human reviewers are financial domain experts.} and performed an iterative annotation for each form component using the DataTorch\footnote{\url{https://datatorch.io/features/annotator}}. Specifically, as shown in Figure~\ref{fig:annotation_procedure}, we split the annotation into three specialized sub-tasks: (1) \textit{Key annotation} for the aforementioned 12 types of \textit{Keys}, (2) \textit{Value annotation} for the paired \textit{Values}, and (3) \textit{Non-Key-Value annotation} for the rest of the non-Key and non-Value components, including \textit{Title}, \textit{Section} and \textit{Others}. Each annotator was assigned one sub-tasks respectively. They identified all possible targeted components and annotated the corresponding rectangular bounding boxes and labels based on the annotation guideline. During the annotation, any uncertain cases would be marked by the annotator and further decided by the discussion and majority voting in the regular meeting. The annotation guidelines were updated whenever needed to handle similar cases later. For instance, one typical type of uncertain case was the annotation of \textit{Form Key} due to its various formats made by different form users, such as \textit{"The previous notice was dated 12 Feb 2003"} as an example for the Key (7) Previous Notice Date, where the final agreed Key \textit{"The previous notice was dated"} is expressed in a complete sentence with the Value \textit{"12 Feb 2023"}. This iterative annotation process produced 757 annotated Digital form documents. In addition, we also randomly selected 50 Printed and 50 Handwritten forms and conducted the annotation process the same as the Digital forms. We included these Printed and Handwritten forms as our additional test set for exploring the possibility of handling different form natures in real-world scenarios.

To ensure the final quality of the human annotation, we visualized the annotated bounding boxes and labels of all forms and assigned them to the two human reviewers for parallel manual-check. They checked each form one by one with reference to the up-to-date annotation guideline and annotated the correctness. They reviewed any form annotated with the incorrect label by both reviewers or received disagreement between them (i.e. one of them annotated as correct while the other annotated as incorrect). Then, those are modified based on the final agreed decision of the two reviewers. To measure the quality of the annotations, we calculated the annotation agreement rate using both Cohen's Kappa \cite{cohen1960coefficient} and the Hamming Loss, which derived the overall scores of \textbf{0.998} and \textbf{0.003}, respectively, indicating a high annotation quality.

\subsection{Annotation Format}
The final annotation of our proposed \textbf{Form-NLU} is provided in \textit{.json} format, separately for Task A and B\footnote{We provide several real examples of the \textit{.json} files for both tasks in \url{https://github.com/adlnlp/form\_nlu\#dataset-loading-and-samples}. The complete dataset will be released upon acceptance of this paper.}. The annotations for each form segment are stored as a \textit{dictionary object} containing multiple key-value pairs with indicative key names in the \textit{.json} file. The main attributes for each form segment shared by the two tasks include \textbf{\textit{bbox}} for bounding box coordinates, \textbf{\textit{text}} for textual tokens of this specific segment (e.g. from OCR or pdfminer\footnote{We use PDFMiner to extract the text of digital-born forms $\mathcal{D}$ and Google Cloud Vision to extract the text of printed and handwritten sets ($\mathcal{P, H}$) }), and the \textbf{\textit{label}} for this segment based on the specific task, e.g., \textit{"Section"} for a segment in Task A or \textit{"Company ID (com\_id)"} for a segment in Task B. Besides, we also include some auxiliary attributes derived from our experiments for potential development in the future, such as the \textbf{\textit{segmentation}} coordinates for Task A as well as the \textbf{\textit{visual\_feature}} and \textbf{\textit{bert\_cls}} for Task B (See Section \ref{sec:taska_model}, \ref{sec:taskb_model}).

\section{Dataset Analysis}\label{sec:dataanalysis}
\subsection{Component Distribution} 
The final version of our annotated~\textbf{Form-NLU} consists of 857 forms, including 757 Digital, 50 Printed and 50 Handwritten forms. Table~\ref{tab:component_distribution} shows the overall statistics of the data splits and the breakdown distribution of form components. We randomly split the Digital forms using the ratio of 70/10/20, resulting in 535, 76 and 146 forms for training, validation and testing. Besides, the 50 Printed and 50 Handwritten forms are included as our additional test splits. The breakdown distribution shows that the \textit{Key} and \textit{Value} persistently dominate since they are the main contents of the form. \textit{Values} can be less than \textit{Keys} due to the cases of empty values. The \textit{Title} and \textit{Section} are the least and demonstrate similar occurrences because each form document normally contains one main title with optional one or two subtitles while always having the two essential sections. The \textit{Others} are slightly more frequent than the \textit{Title} or \textit{Section}, and the occurrence depends on the different form fillers. This consistent overall distribution across forms is attributed to the use of the Form 604 template. It makes our \textbf{Form-NLU} a promising benchmark dataset for exploring solutions to the real-world financial form layout understanding problem when the standard form template is available (our Task A). In addition, we further differentiate the keys and values for comprehending the specific form intentions (our Task B). These key-value pairs are contained in either \textit{Table} (mostly vertical key-value alignment) or \textit{Form} (mostly horizontal key-value alignment), which indicates the variety of spatial relationships that requires the model to learn in order to accurately identify the values and align with each key. In Table~\ref{tab:bbox_info}, we provide the average bounding box size and textual tokens of each form component, which reflects both their visual and linguistic features that can be potentially helpful for specific key and value identification. \textit{Title} and \textit{Section} tend to have longer textual content and bigger component size with larger font sizes. In comparison, \textit{Key} and \textit{Value} are much shorter and are contained in smaller bounding boxes. Especially the \textit{Value} in the \textit{Table} has the shortest length as they are mostly numeric values.

\begin{table}[t]
    \begin{center}
    \begin{adjustbox}{max width = \linewidth}
    \begin{tabular}{l|l|l|lllllll}
    \hline
     \multirow{2}{*}{\textbf{Type}}& \multirow{2}{*}{\textbf{Usage}}&\textbf{Form}& \multirow{2}{*}{\textbf{Title}} & \multirow{2}{*}{\textbf{Section}}  & \multicolumn{2}{c}{\textbf{Form}} & \multicolumn{2}{c}{\textbf{Table}} & \multirow{2}{*}{\textbf{Others}}\\
     \cline{6-9}
    &&\bf Images& &   & \textbf{Key} & \textbf{Value} & \textbf{Key} &\textbf{Value} &\\
    
    \hline
    \textbf{Train} & \multirow{2}{*}{\textbf{Digital}} & 535 &1068 &1070  &3708 &3568  &2669 &2669 &1691 \\
    \cline{1-1}
    \cline{3-10}
    \textbf{Val}& & 76 & 152 &152  &524 &510  &380 &379 &246\\
    \cline{1-1}
    \cline{2-10}
    \multirow{3}{*}{\textbf{Test}}& \bf Digital & 146 & 292 &292  &1009 &978  &730 &730 &458\\
    \cline{2-10}
    &\textbf{Printed} & 50 & 98 &100  &346 &332  &250 &249 &152\\
    \cline{2-10}
    &\textbf{Handwritten}& 50 & 100 &100  &348 &315  &249 &226 &149\\
    \hline
    \multicolumn{2}{l|}{\bf Total Number} & 857 & 1710 & 1714 & 5935 & 5703 & 4278 & 4253 & 2696 \\
    \hline
    \end{tabular}
    \end{adjustbox}
    \end{center}
    \caption{Number of Form Components for Each Data Split.}
    \label{tab:component_distribution}
    \vspace*{-1.5em}
\end{table}
\begin{table}[t]
    \begin{center}
    \begin{adjustbox}{max width =\linewidth}
    \begin{tabular}{l|llllll}
    \hline
     \multirow{2}{*}{\textbf{Type}}&\multirow{2}{*}{\textbf{Title}} & \multirow{2}{*}{\textbf{Section}}  & \multicolumn{2}{c}{\textbf{Form}} & \multicolumn{2}{c}{\textbf{Table}}\\
     \cline{4-7}
    &&&     \textbf{Key} & \textbf{Value} & \textbf{Key} &\textbf{Value}\\
    \hline
    \textbf{Bounding Box Average Width}  &186.15 &136.82   &95.46 &85.98   &57.29 &41.92\\
    \textbf{Bounding Box Average Height} &30.44 &12.47   &12.47 &12.47   &10.50 &10.45\\
    \textbf{Bounding Box Average PX}   &4275.94 &1720.62   &1204.96 &1183.10   &609.75 &456.94\\
    \textbf{Average Number of Tokens}  &7.35 &7.10 &5.16 &4.14 &4.29 &1.74\\
    \hline
    
    \end{tabular}
    \end{adjustbox}
    \end{center}
    \caption{Average Bounding Box Width, Height, Number of Pixel(PX), and Number of Tokens for Each Form Component}
    \label{tab:bbox_info}
    \vspace*{-2em}
\end{table}

\begin{figure*}[t]
    \hspace*{-0.3em}
    
     \centering
     \begin{subfigure}[b]{0.15\textwidth}
         \centering
         \includegraphics[height=3.3cm]{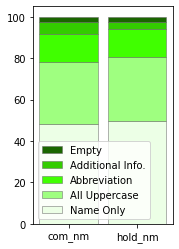}
         \caption{~~name}
         \label{fig:name}
     \end{subfigure}
     \hspace*{-0.3em}
     \begin{subfigure}[b]{0.15\textwidth}
         \centering
         \includegraphics[height=3.3cm]{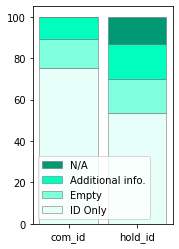}
         \caption{~~id}
         \label{fig:id}
     \end{subfigure}
     \hspace*{-0.3em}
     \begin{subfigure}[b]{0.225\textwidth}
         \centering
         \includegraphics[height=3.3cm]{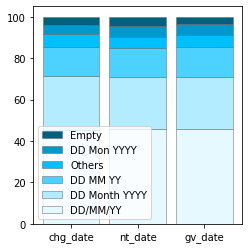} 
         \caption{~~dates}
         \label{fig:date}
     \end{subfigure}
     \hspace*{-0.5em}
     \begin{subfigure}[b]{0.08\textwidth}
         \centering
         \includegraphics[height=3.3cm]{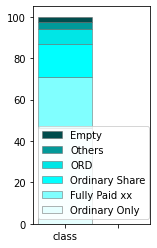}
         \caption{~~class}
         \label{fig:class}
     \end{subfigure}
     \hspace*{1.3em}
     \begin{subfigure}[b]{0.32\textwidth}
         \centering
         \includegraphics[height=3.3cm]{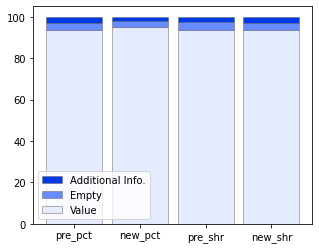}
         \caption{~~percent/share}
         \label{fig:percent/share}
     \end{subfigure}
        \caption{Value Patterns Distributions for Each Key Group.}
        \label{fig:key value pattern}
\end{figure*}

\subsection{Value Pattern Proportion} 
As mentioned before, the filled-in content varies and involves potential noises due to the diversified user understanding and uncertainty. To illustrate the variety of the filled-in content, we summarise the main value patterns for each key intention and provide the proportions in Figure~\ref{fig:key value pattern}. Throughout all keys, the largest proportion of value follows the common value pattern that the form users tend to provide minimal information to fulfil the intention of the keys. For example, they put simply the company name only to the key \textit{com\_nm} in Figure~\ref{fig:name} or ID only for \ref{fig:id}. 
However, there are several other value patterns available, including additional information shown in Figure~\ref{fig:name}, \ref{fig:id} and \ref{fig:percent/share}, or having different data format or writing style, such as the various date formats in Figure~\ref{fig:date}. These human-caused variants well reflect the real-world form understanding scenarios and imply potential challenges for achieving precise information retrieval in our Task B. Overall, the Share Class (Figure~\ref{fig:class}) and the date type values (Figure~\ref{fig:date}) show comparatively more patterns while the numerical type values in Figure~\ref{fig:percent/share} tend to be more consistent with the common pattern. This pattern similarity among the keys with the same data type values indicates that solely relying on the linguistic cues may not be enough as understanding the relative spatial location of the values in the form is also required in order to distinguish these similar values of different keys.   

\subsection{Key-value Pair Comparison Analysis} 
\begin{figure}[t]
    \hspace*{0cm}
     \centering
     \hspace*{-0.5em}
     \begin{subfigure}[b]{0.235\textwidth}
         \centering
         \includegraphics[height=2.5cm]{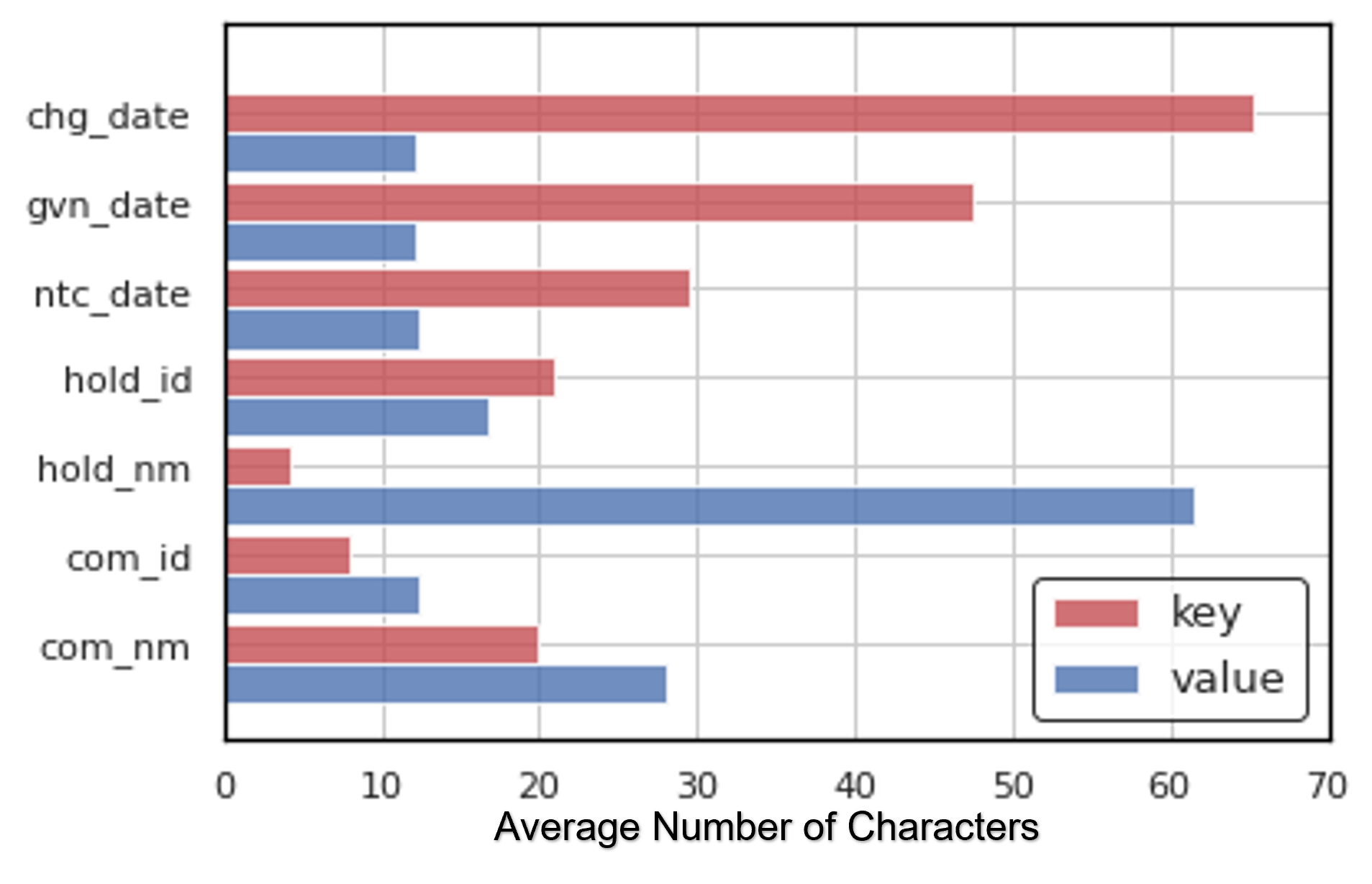}
         \caption{Form Key Value Pairs}
         \label{fig:form char number}
     \end{subfigure}
     \hspace*{-0.5 em}
     \begin{subfigure}[b]{0.235\textwidth}
         \centering
         \includegraphics[height=2.5cm]{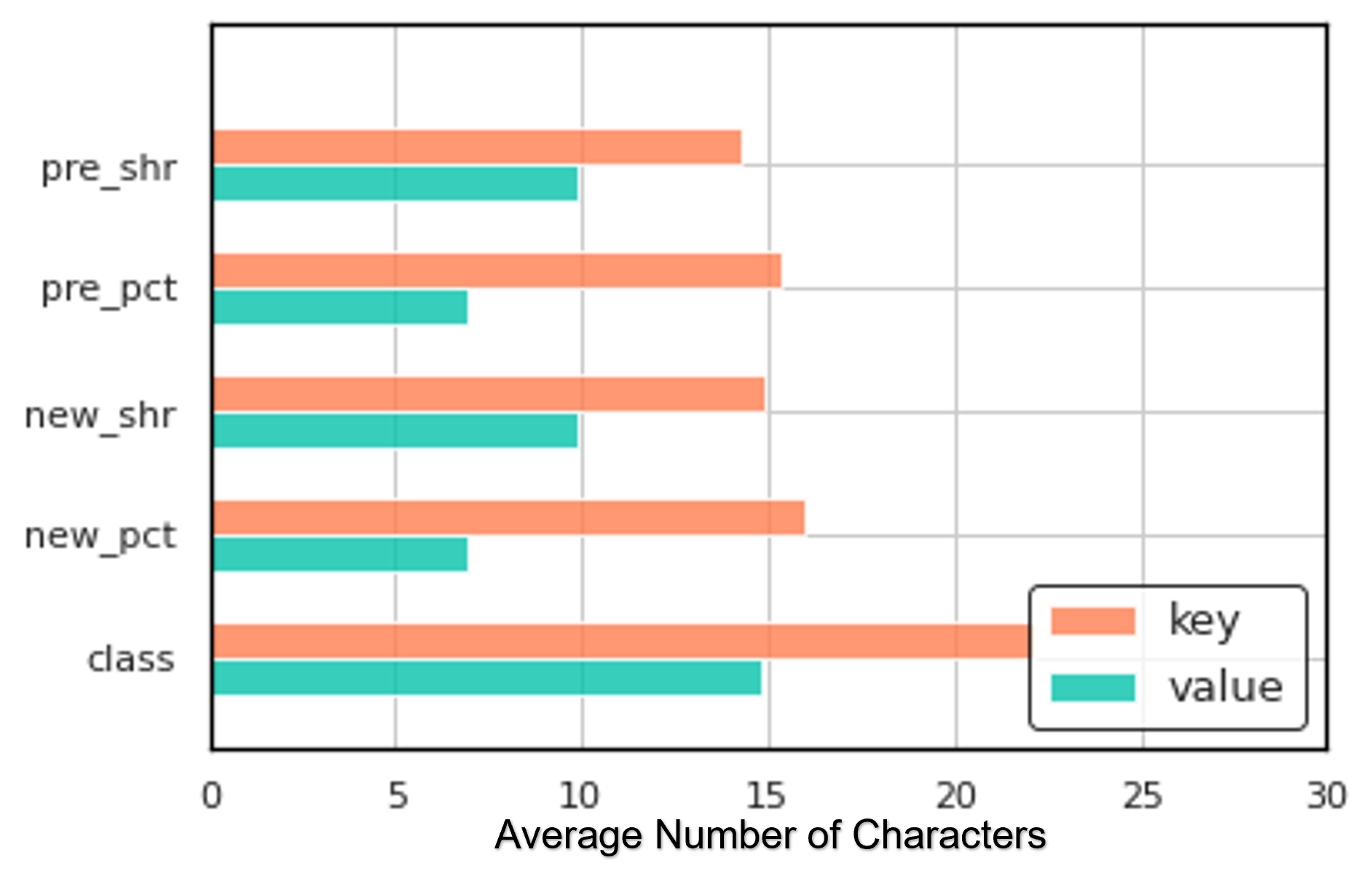}
         \caption{Table Key Value Pairs}
         \label{fig:table char number}
     \end{subfigure}
        \caption{Average \# of Characters in Each Key Value Pair}
        \label{fig:character}
        \vspace*{-1em}
\end{figure}

\begin{figure}[t]
    \hspace*{-0.5em}
     \begin{subfigure}[b]{0.24\textwidth}
         \centering
         \includegraphics[height=2.5cm]{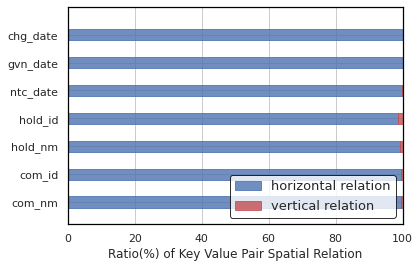}
         \caption{Form Key Value Pairs}
         \label{fig:form spatial ratio}
     \end{subfigure}
     \hspace*{-0.5em}
     \begin{subfigure}[b]{0.24\textwidth}
         \centering
         \includegraphics[height=2.5cm]{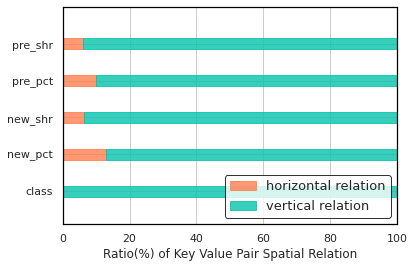}
         \caption{Table Key Value Pairs}
         \label{fig:table spatial ratio}
     \end{subfigure}
        \caption{Ratio (\%) of Horizontal and Vertical Relation for Each Key Value Pair}
        \label{fig:relation}
        \vspace*{-1.5em}
\end{figure}

To provide an in-depth analysis of the nature of key-value pairs, we further summarize the average character number and the ratio of spatial relation for the twelve key-value pairs in Figure~\ref{fig:character} and~\ref{fig:relation}, respectively. It can be observed that typically there are seven fixed key-value pairs in Form (i.e., Figure~\ref{fig:form char number}/\ref{fig:form spatial ratio}) and another fixed five in Table (i.e., Figure~\ref{fig:table char number}/\ref{fig:table spatial ratio}) formatted by the standard Form 604 template. Two groups of key-value pairs demonstrate their own features. As shown in Figure~\ref{fig:form char number} for the form-based key-value pairs, the three date type pairs tend to have much longer value than the key where as the other four pairs may have either longer or shorter values than keys. The \textit{hold\_nm} key-value shows the largest character length gap among all. In comparison, the five Table-based pairs in Figure~\ref{fig:table char number} all have shorter values than keys, and the length gaps are similar. Overall, most keys have short values (e.g., $< 20$ characters) from which the semantic context could be scarce. On the other hand, as seen from Figure~\ref{fig:relation}, the seven Form-based pairs are uniformly aligned horizontally, whereas the other five pairs in Table are mostly vertical. However, some special cases with the opposite alignment are also observed in both groups, such as the vertical key-value pairs for the \textit{hold\_id}/\textit{hold\_nm} and \textit{com\_id}/\textit{com\_nm} in the Form-based group (Figure~\ref{fig:form spatial ratio}), as well as the horizontal pairs for the four percentage type keys and values in the Table-based group (Figure~\ref{fig:table spatial ratio}). Thus, simply memorizing the spatial alignment for each key cannot lead to optimized value retrieval in Task B.    


\section{Tasks Overview}\label{sec:task_overview}
\begin{figure}[t]
    \hspace*{-0.5em}
     \centering
     \begin{subfigure}[b]{0.24\textwidth}
         \centering
         \includegraphics[height=4.1cm]{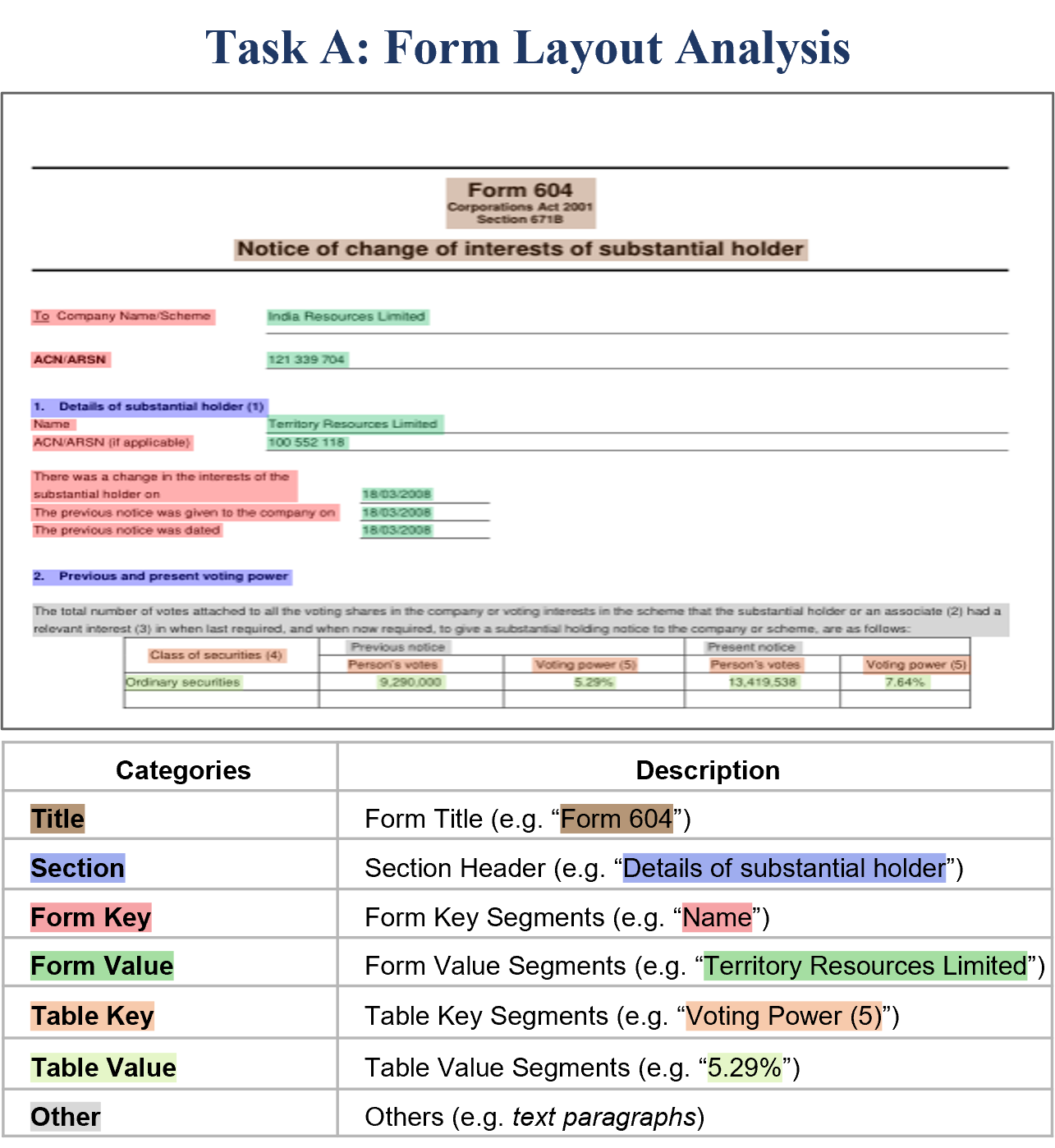}
         \caption{Task A Sample}
         \label{fig:taska_definition}
     \end{subfigure}
     \hspace*{-0.5em}
     \begin{subfigure}[b]{0.24\textwidth}
         \centering
         \includegraphics[height=4.1cm]{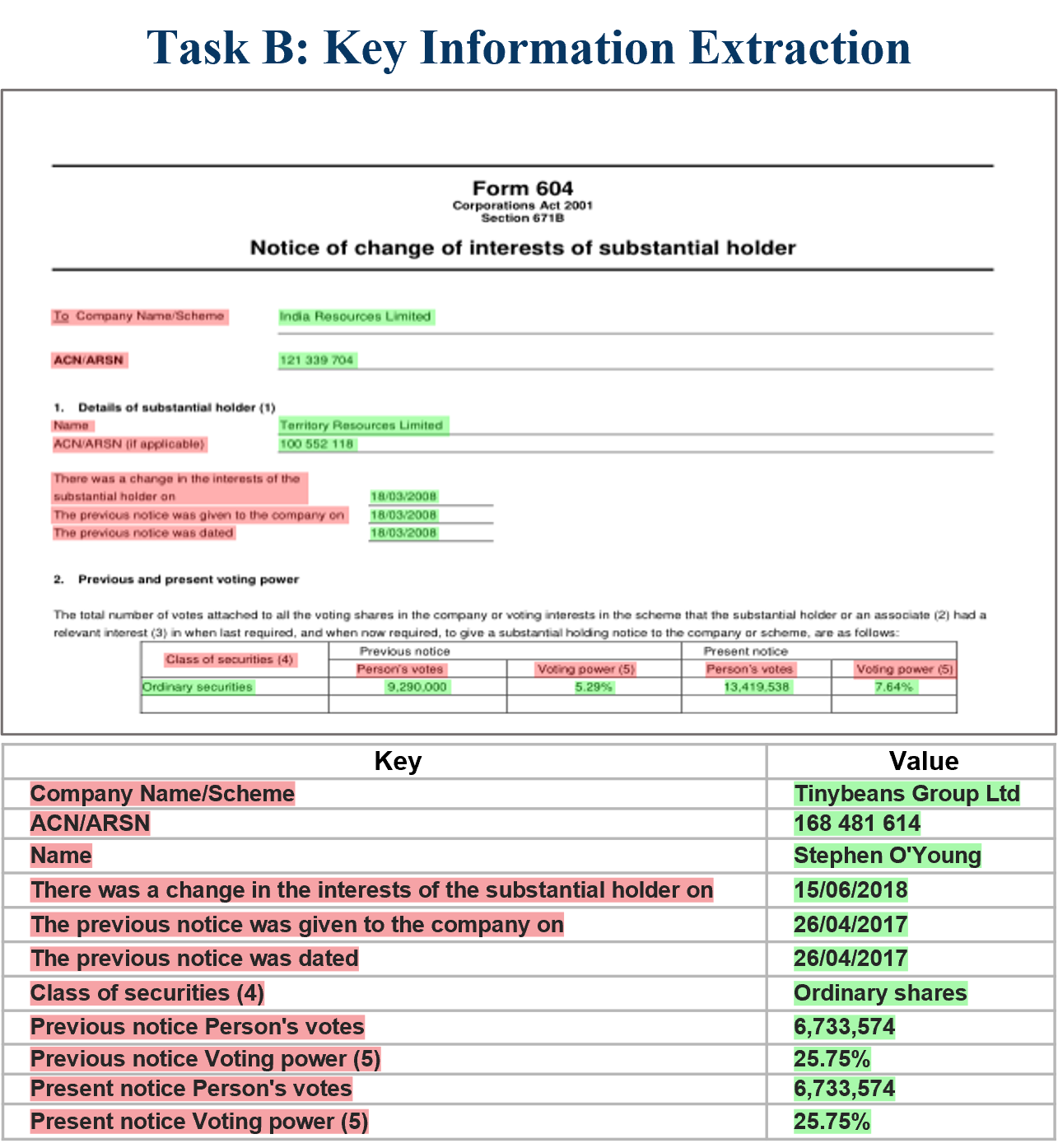}
         \caption{Task B Sample}
         \label{fig:taskb_definition}
     \end{subfigure}
        \caption{Examples for Task A and B. For Task A, the users need to detect each layout component's bounding box and recognize the associated category. Task B asks the user to feed the fixed key text (red area) into the model to predict the corresponding value of RoI's index (green area)}
        \label{fig:task_definition}
         \vspace*{-1em}
\end{figure}
\subsection{Task A - Form Layout Analysing}\label{sec:taskAdef}
The purpose of Task A is to detect the semantic entities (\textit{Title}, \textit{Section}, \textit{Form\_key}, \textit{Form\_value}, \textit{Table\_key},  \textit{Table\_value}, \textit{Others}) of forms. This is a prior task to understand the form layout by detecting the position of each semantic entity, including key, value and other objects. Given a form image $I$, an object detection model is used to detect a set of RoIs $(r_1,r_2,...,r_n)$. Each $r_i$ contains bounding box $b_i$ coordinates $(x_i,y_i,w_i,h_i)$ with semantic category $c_i$. As shown in Figure~\ref{fig:taska_definition}, the bounding box of each semantic entity in an input form image is detected and coloured based on the recognized categories. The model should effectively detect layout components such as form \textit{Title}  (like "\colorbox{brown!30}{Form 604}") and \textit{Section} headers (such as "\colorbox{blue!30}{Details of substantial holder}"). Additionally, we also expect the models can differentiate keys/values located in \textit{Form} or \textit{Table}; for example, "\colorbox{red!30}{Name}" should be detected as a \textit{Form\_key} instance, while "\colorbox{orange!30}{Voting Power (5)}" is a \textit{Table\_key} instance. 

\subsection{Task B - Key Information Extraction}\label{sec:taskBdef}
Task B aims to evaluate whether the proposed models could comprehensively understand designers' intentions and user uncertainties to extract valuable information from input forms. It allows using ground truth RoIs' set $\mathcal{R}_{gt} = (R_1,R_2,...,R_n)$ during the training and inference stage. Given key text information $t$, a document image $I$ and a set of ground truth RoIs $(r_1,r_2,...,r_n)$, a model $H$ can output the RoI's index number aligned with input $t$. As Figure~\ref{fig:taskb_definition} shown, each highlighted entity with red background colour is the key $t$ we need to feed into the proposed model, and the green paired RoI's index is the desired output from the model based on the current input key. For example, inputting the required key text content "\colorbox{red!30}{Company Name/Scheme}", the desired output from the model is the RoI's index of paired values where we can easily get the text content ("\colorbox{green!30}{Tinybeans Group Ltd}") from it.



\section{Evaluation Setup}
\subsection{Implementation Detail}
For Task A, we fine-tune Faster-RCNN and Mask-RCNN models with two backbones\footnote{Faster-RCNN backbones are faster\_rcnn\_R\_50\_FPN\_3x, faster\_rcnn\_R\_101\_FPN\_3x and Mask-RCNNs are mask\_rcnn\_R\_50\_FPN\_3x, mask\_rcnn\_R\_101\_FPN\_3x} respectively on our dataset based on Detectron2 platform. We set 5000, 128, and 0.02 as the maximum iteration times, batch size and base learning rate, and other setups are the same as Detectron2 official tutorial \footnote{\url{https://colab.research.google.com/drive/16jcaJoc6bCFAQ96jDe2HwtXj7BMD_-m5}}. Regarding Task B, we employ various approaches to encode the vision and language features. Firstly, all Task B adopted baselines use pretrained BERT to encode key textual content. Moreover, for the visual aspect, VisualBERT, LXMERT, and M4C models utilize 2048-d features extracted from the Res5 layer of ResNet101. The maximum number of input key text tokens and the number of segments on each page are all defined as 50 and 41, respectively. Task A and B experiments are conducted on 51 GB Tesla V100-SXM2 with CUDA11.2.

\subsection{Task A Baselines and Metric}
We use two popular object detection models, \textbf{FasterRCNN}~\cite{fasterrcnn} and \textbf{MaskRCNN}~ \cite{maskrcnn} with different ImageNet pretrained backbones testing on our dataset. 
To evaluate the performance of the object detection model, we apply a $mAP$ (mean Average Precision), which is commonly used in object detection tasks. The $mAP$ score is calculated by averaging $APs$ over all categories' overall pre-defined IoU thresholds\footnote{We refer \cite{publaynet} to adopt $mAP$ as metrics and follow their thresholds for Task A.}.

\subsection{Task B Baselines and Metric}
Task B mainly focus on predicting the corresponding value of the RoI index based on input text content (as shown in Figure~\ref{fig:taskb_definition}). Thus, several multi-modal transformer frameworks are adopted as baselines on \textbf{Form-NLU}, of which inputs are multi-aspect RoI features, including large pre-trained \textbf{VisualBERT}~\cite{visualbert}, \textbf{LXMERT}~\cite{tan2019lxmert} and non-pre-trained \textbf{M4C}~\cite{m4c} models \footnote{Detailed baseline setup can be found in \url{https://github.com/adlnlp/form\_nlu\#baseline-model-description}}. Regarding evaluation, we use weighted F1-score as the primary evaluation metrics for representing overall and breakdown performance.
Note that some visual language pre-trained models, such as ViLT~\cite{kim2021vilt}, LayoutLM~\cite{xu2020layoutlm}, were excluded since those are mainly based on the image patches or pieces and do not fit into the task demands.

\subsection{Task A Model}
\label{sec:taska_model}
For Task A - Form Layout Analysing task (Section ~\ref{sec:taska_results}), we use Faster-RCNN and Mask-RCNN with various depth backbones (ResNet-50 and ResNet-101) as baselines and contact experiments to check the effects of diverse model architecture and model size. Additionally, for the real-world case analysis (Section~\ref{sec:case_study1}), we adopted some methods using open-source PDF parsers (such as PDFMiner) to analyse the form layout without training the deep learning models. The PDF parser outputs can also be used as inputs for Task B.
\subsection{Task B Model}
\label{sec:taskb_model}
\begin{figure}[t]
     \centering
         \includegraphics[width=0.45\textwidth]{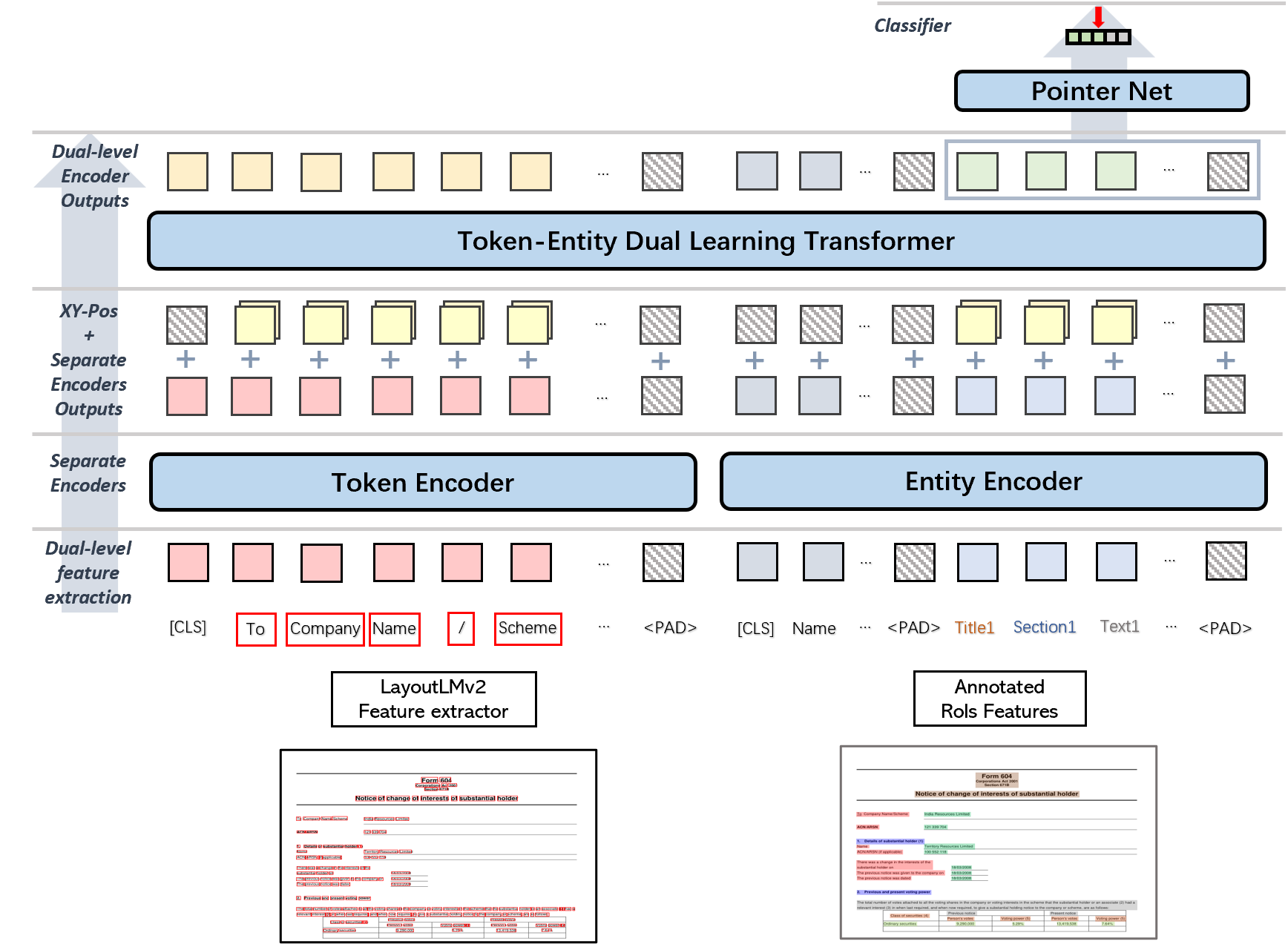}
         \caption{Our Proposed Key Information Extraction Model Architecture for Task B}
         \label{fig:model architecture}
         \vspace*{-1em}
\end{figure}

For Task B, we propose a new document key information extraction model (as Figure~\ref{fig:model architecture} shown) to predict the corresponding values from input key content. To achieve this, we investigate positional and logical relations and their hierarchical structure with two components: multiple aspect features extraction/integration and entity-token dual-level Model. 
\subsubsection{Multi-Aspect Features}
We investigate five aspect features, including Visual ($V$), Textual ($T$), Positional ($P$), Density ($D$), and Gap Distance ($G$) features, with the corresponding encoding approaches to comprehensively explore which features benefit understanding designer intentions and argument form-like document segment representations. Many traditional methods for document understanding tasks have shown the effectiveness and significance of visual, textual, and positional features only \cite{xu2020layoutlm, xu2021layoutlmv2, zhang2021entity}. Except for those three aspect features, we also aim to explore the effectiveness of text density for form understanding, which has been demonstrated by \cite{docgcn} for document layout analysis. Moreover, based on the dataset analysis results, the gap distance between entities is crucial for understanding the form structure. Hence, we introduce two layout-related features, including normalized positional features (bounding box coordinates) and the gap distance between an entity to neighbours
\footnote{Our Form-NLU Github (\url{https://github.com/adlnlp/form\_nlu\#multi-aspect-features}) provides more detailed feature representation approaches.}.

\subsubsection{Entity-Token Dual Level Model}
The proposed new entity-token dual-level form information extraction model contains four components, including Entity Encoder $Encoder_{entity}$, Token Encoder $Encoder_{token}$, Dual Encoder $Encoder_{dual}$ and Pointer Net-based classifier $\phi$.

\textbf{1) Entity Encoder} aims to learn the semantic relations between entities for enhancing the vanilla entity representations. Based on the preliminary results, we select the pretrained LXMERT \ to cite{tan2019lxmert} as the entity encoder. The inputs of the LXMERT-based entity encoder include key text $T_{key}$, Entities' visual features $V_{entity}$, and the bounding box coordinates $B_{entity}$. After feeding those features into pretrained LXMERT, the enhanced Entity RoIs' representations $E \in \mathbb{R}^{768}$ can be extracted. 

\textbf{2) Token Encoder} aims to acquire cross-modal token representations. The SoTA layout-aware pre-trained visual language transformer, LayoutLMv2~\cite{xu2021layoutlmv2}, is adopted as the token encoder. We firstly employ \textit{LayoutLMV2FeatureExtractor} to extract and encode the token level features, including token text $T_{token}$, token RoIs' bounding box $B_{token}$ with pixel features $I_{pixel}$. Then, we follow the original LayoutLMV2 setup to process the extracted token-level features and feed them into our token-level encoder $Encoder_{token}$, pretrained LayoutLMv2 to get the token-level representation $T \in \mathbb{R}^{768}$. 

\textbf{3) XY-Pos Dual Encoder} is designed for learning the geometrically sensitive multi-grained and multi-modality feature representations. 
After getting entity and token level feature representations from pretrained models, we treat them as the sequence inputs of a dual-level mutual learning encoder $Encoder_{dual}$, where a 6-layer transformer with 8-\textit{heads} self-attention is used as the basic framework. Unlike the original transformer that adopts sinusoidal positional encoding \cite{transformer}, we proposed a new geometric-sensitive positional encoding \textit{XY-Pos}. It flattens the stacked sets of normalized bounding box coordinates along with the X or Y axis to get $X_i^{pos}$ and $Y_i^{pos}$. Supposing $b_i = (x_i, y_i, w_i, h_i)$ is the bounding box of $r_i$ of $Encoder_{dual}$ in document page $D_j$, the size of $D_j$ is $(W_j, H_j)$. 
\begin{equation}
    pos_{i}^x = [\frac{x_i+d_{i_1}}{W_j}, \frac{x_i+d_{i_2}}{W_j},...,\frac{x_i+d_{i_m}}{W_j}]
\end{equation}
where the $l$-th step $d_{i_l} =  \frac{w_i\times l}{m}$
\begin{equation}
    X_i^{pos}=flatten([pos_{i}^x] \times n)
\end{equation}
 We set $m$ and $n$ as 32 and 24 to get a 768-d vector. Finally $X_i^{pos}$ is flattened into a $m\times n$ dimensional vector, we have $X_i^{pos} \in \mathbb{R}^{768}$. Similar procedures are used to generate positional encoding $Y_i^{pos}$  along \textit{Y-axis} of $r_i$. The final input representations before feeding into $Encoder_{dual}$ can be represented as:
\begin{equation}
    E_{pos_{xy}} = E+X_{entity}^{pos}+Y_{entity}^{pos}
\end{equation}
\begin{equation}
    T_{pos_{xy}} = T+X_{token}^{pos}+Y_{token}^{pos}
\end{equation}

Unlike most VLPMs \cite{tan2019lxmert, xu2021layoutlmv2} adopted positional encoding methods through linear projecting 4-d bounding box coordinates into high dimensional vectors. \textit{XY-pos} could capture more geometric features between entities and tokens. Then we will feed them into $Encoder_{dual}$ and get the updated token and entity representations $T_{dual}$, $E_{dual}$.

\textbf{4) Pointer Net based Classifier}
The multi-aspect features are concatenated with $E_{dual}$ to get $E_{multi}$. $E_{multi}$ is fed into the pointer net \cite{vinyals2015pointer} based classifier $\phi$ to get a score vector which following a $softmax$ to retrieve the final prediction results $y_{pre}$. 

\section{Results}
\subsection {Task A: Form Layout Analysing}
\label{sec:taska_results}
\subsubsection{Overall and Breakdown Performance}
\begin{table}[t]
    \begin{center}
    \begin{adjustbox}{max width =\linewidth}
    \begin{tabular}{c|l|l|lllllll}
    \hline
    \multirow{2}{*}{\textbf{Test set}} & \multirow{2}{*}{\textbf{Model}} & \multirow{2}{*}{$mAP$}  & \multicolumn{7}{c}{\textbf{Breakdowns ($Precision$)}}\\
    \cline{4-10}
    &  &  & \textbf{Title} & \textbf{Section} & \textbf{Form\_key} &\textbf{Form\_value} & \textbf{Table\_key} &\textbf{Table\_value} & \textbf{Others}\\
     \hline
     
    \multirow{4}{*}{\shortstack{$\mathcal{D}$}}&\textbf{F-50} & 68.98& 73.54& 69.36& 67.93& 70.37& \textbf{69.18}& 66.76&65.73\\
    &\textbf{F-101} & 69.99& 71.32& 68.30& 72.37& 71.70& 68.74&72.84& 64.65\\
    &\textbf{M-50} & 71.74& 77.29& 69.06& \textbf{73.01}& 70.30& 68.40& 73.25&\textbf{70.90}\\
    &\textbf{M-101} & \textbf{71.93}& \textbf{79.54}& \textbf{69.64}& 71.25& \textbf{72.38} & 67.02& \textbf{74.38}&69.33\\
    \hline
    \multirow{4}{*}{\shortstack{$\mathcal{P}$}}&\textbf{F-50}& 56.46& 57.44& 53.98& 54.49& 51.46& 59.31& 62.86& 55.67\\
    &\textbf{F-101} &59.54& 55.72& 52.81& 63.13& 61.32& 59.42& 68.11& 56.26\\
    &\textbf{M-50}&63.06 &59.83 &60.98 &63.74 &\textbf{61.91} &70.75 &65.93 &58.29\\
    &\textbf{M-101} &\textbf{65.47} &\textbf{63.05} &\textbf{62.49} &\textbf{66.84} &61.80 &\textbf{71.82} &\textbf{70.19} &\textbf{62.10}\\
    \hline
    \multirow{4}{*}{\shortstack{$\mathcal{H}$}}&\textbf{F-50}&49.87 & 58.73&43.85 &63.14 &25.15 &63.48 &40.14 &54.56\\
    &\textbf{F-101}& 50.39& 53.92& 33.36& 65.28&36.68&66.97 &41.06 &55.43\\
    &\textbf{M-50} & 57.99& 61.45& 48.46& 68.16&43.14& 71.76&\textbf{56.95} &56.05\\
    &\textbf{M-101}& \textbf{60.22}& \textbf{64.81}& \textbf{54.46}& \textbf{69.97}& \textbf{46.82}& \textbf{73.34}&52.81 &\textbf{59.32}\\
    \hline
    \end{tabular}
    \end{adjustbox}
    \end{center}
    \caption{Overall Performance and Breakdown Results for Layout Analysing Task (Task A). F-50 and F-100 represent the Faster-RCNN with ResNet50 and ResNet101 as backbones. The same patterns occur for Mask-RCNN (M-50 or M-101). }
    
    \label{tab:taska}
    \vspace*{-2.5em}
\end{table}

This section shows the test performance of layout analysing (Task A) models on digital ($\mathcal{D}$), printed ($\mathcal{P}$), and handwritten ($\mathcal{H}$) sets. From Table~\ref{tab:taska}, we observe that the $mAP$ of Mask-RCNNs can achieve around 2\%, 6\% and 10\% higher than Faster-RCNNs with identical backbones on $\mathcal{D}$, $\mathcal{P}$ and $\mathcal{H}$, respectively. This may result from the finer spatial localisation of auxiliary instance segmentation tasks adopted by Mask-RCNNs. Then, we explore the effects of various backbones, like Faster-RCNN with ResNet-101 (F-101) or Mask-RCNN with ResNet-50 (M-50). From Table~\ref{tab:taska}, we can find F-101 and M-101 consistently achieve around 1\% to 2\% higher than F-50 and M-50. It demonstrates that the deeper or large-scale pretrained backbones may generate more comprehensive visual representations. Notably, the overall performance of $\mathcal{D}$ is around 5\% higher than $\mathcal{P}$ and even about 11\% more than $\mathcal{H}$. The main reason may result from the apparent difference between scanned (both printed $\mathcal{P}$ and handwritten $\mathcal{H}$) and digital ($\mathcal{D}$) forms, especially the handwritten forms involving more user-uncertainties such as writing mistakes or scanning rotation. 

As our Form-NLU provides fine-grained layout component types such as \textit{Value} subdivided into \textit{Form\_value} and \textit{Table\_value}, it enables us to observe and analyse the performance of subdivided components affected by distinct layout position distribution.  For example, certain layout components show stable performance among the three test sets, such as \textit{Form\_key} and \textit{Table\_key}. It may result from those components designed by form designers with more shared layout and visual patterns like font and layout arrangement. However, \textit{Form\_value} and \textit{Table\_value} components may involve more uncertainties and noise from users, electric devices or propagation process, leading to $mAP$ on $\mathcal{H}$ (46.82\% and 52.81\%) being lower than $\mathcal{P}$ (61.80\% and 70.19\%) and much lower than $\mathcal{D}$ (72.38\% and 74.38\%). In addition, for \textit{Title} and \textit{Others} components, the performance of M-101 on $\mathcal{D}$ (79.54\% and 69.33\%)  is apparently higher than $\mathcal{P}$ (63.05\% and 62.10\%) and $\mathcal{H}$ (64.81\% and 59.32), while $\mathcal{P}$ and $\mathcal{H}$ almost have similar performance. The reason may come from the difference between scanned ($\mathcal{P}$ and $\mathcal{H}$) and digital forms, such as the rotation or lower resolution of scanned forms.

\subsubsection{Stepped Training Set Ratios}
\begin{figure}[tb]
    \hspace*{0cm}
     \centering
     \hspace*{-1.9em}
     \begin{subfigure}[b]{0.13\textwidth}
         \centering
         \includegraphics[height=2.3cm]{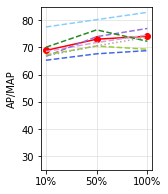}
         \caption{Digital($\mathcal{D}$)}
         \label{fig:taska_test}
     \end{subfigure}
     \hspace*{-1em}
     \begin{subfigure}[b]{0.13\textwidth}
     
         \centering
         \includegraphics[height=2.3cm]{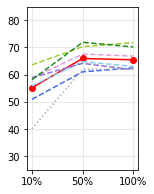}
         \caption{Printed($\mathcal{P}$)}
         \label{fig:taska_printed}
     \end{subfigure}
     \hspace*{-0.5em}
     \begin{subfigure}[b]{0.18\textwidth}
     \captionsetup{justification=raggedright,singlelinecheck=false}

         \centering
         \includegraphics[height=2.3cm]{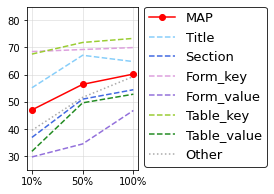}
         \caption{Handwritten($\mathcal{H}$)}
         \label{fig:taska_handwritten}
     \end{subfigure}
        \caption{Different Training Ratio $mAP$ on Digital ($\mathcal{D}$), Printed ($\mathcal{P}$), and Handwritten ($\mathcal{H}$) Test Sets for Task A}
        \label{fig:taska_ratio}
        \vspace*{-1.5em}
\end{figure}

We set stepped ratios of training set size ($\mathcal{T}$) (10\%, 50\% and 100\%) to train M-101 and evaluate it on $\mathcal{D}$, $\mathcal{P}$, and $\mathcal{H}$ sets to explore the effects of training size on Task A. It can be seen from Figure~\ref{fig:taska_ratio} as $\mathcal{T}$ size increases, overall and breakdown performance improve, whereas this trend becomes inapparent when $\mathcal{T}$ reaches 50\% on $\mathcal{D}$ and $\mathcal{P}$. Regarding performance on $\mathcal{H}$, increasing the training size could boost the performance of all types of layout components, significantly for \textit{Table\_value}, \textit{Section} and \textit{Others}. Especially, as one significant layout component type for downstream tasks, \textit{Table\_value} hold around 30\% on $\mathcal{H}$. In general, the trends shown in Figure~\ref{fig:taska_ratio} demonstrate that our training set ($\mathcal{T}$) contains various form types to enable handling distinct application scenarios, even with the small ratio of $\mathcal{T}$. Moreover, it also represents increasing the size of digital training sets ($\mathcal{T}$) that could effectively improve printed and handwritten performance. More training samples may enable catching more shared feature patterns, such as spatial distributions, typographical similarity, font type and size of specific components.

\subsection {Task B: Key Information Extraction} 
\label{sec:taskb_results}
\subsubsection{Overall and Multi-aspect Feature Performance}

\label{sec:taskb_results_overall_performance}
This section compares and analyses the performance between our model and three widely used baselines under different configurations on digital ($\mathcal{D}$), printed ($\mathcal{P}$) and handwritten ($\mathcal{H}$) test sets. Firstly, we focus on comparing the performance among vanilla models (the first row of each model group in Table~\ref{tab:taskb1_aspects}) in which input features and model architectures are the same as their paper described. Our model achieves better performance than all vanilla baselines on $\mathcal{D}$, $\mathcal{P}$ and $\mathcal{H}$. It indicates that iterative learning with token-level features could enhance entity-level representations for boosting downstream performance. In addition, for three vanilla baselines, LXMERT can get 2.28\%, 6.38\% and 15.36\% higher than VisualBERT on $\mathcal{D}$, $\mathcal{P}$ and $\mathcal{H}$, respectively, which may result from inputs of LXMERT containing positional information, but VisualBERT is pre-trained on the visual feature of input RoIs only. Subsequently, compared with VisualBERT, the non-pretrained M4C can increase by around 1.5\%, 3\% and 9\% on $\mathcal{D}$, $\mathcal{P}$ and $\mathcal{H}$, respectively. It may illustrate the significance of textual and positional features. However, due to the input feature differences between vanilla models, we conducted external experiments to explore the effects of multi-aspect features on all adopted models for a fair comparison.

\begin{table}[t]
    \begin{center}
    \begin{adjustbox}{max width = 0.9\linewidth}
    \begin{tabular}{c|p{0.015\textwidth}p{0.015\textwidth}p{0.015\textwidth}|p{0.015\textwidth}p{0.015\textwidth}|p{0.08\textwidth}|p{0.08\textwidth}|p{0.11\textwidth}}
    \hline
    \multirow{2}{*}{\textbf{Model}} &\multicolumn{5}{c|}{\textbf{Input Features}} & \multicolumn{3}{c}{\textbf{Overall (F1-score)}}\\
    \cline{2-9}
     &\textbf{$V$} &\textbf{$T$}&\textbf{$P$}& \textbf{$D$} & \textbf{$G$}  & Digital(\textbf{$\mathcal{D}$}) & Printed(\textbf{$\mathcal{P}$}) & Handwritten(\textbf{$\mathcal{H}$})\\
    \hline 
    \multirow{4}{*}{\textbf{M4C}} & \cellcolor{orange!20} $\bigcirc$ & \cellcolor{orange!20} $\bigcirc$ & \cellcolor{orange!20} $\bigcirc$ & \cellcolor{orange!20} $\times$ & \cellcolor{orange!20} $\times$ & \cellcolor{orange!20} 96.91 & \cellcolor{orange!20} 88.62 & \cellcolor{orange!20} 74.06   \\
    \cline{2-9}
     & $\bigcirc$ & $\bigcirc$ & $\bigcirc$ & $\bigcirc$ & $\times$ & 96.32 &89.52 &74.98   \\
     & $\bigcirc$ & $\bigcirc$ & $\bigcirc$ &$\times$ & $\bigcirc$ & 97.00 &88.81 & \underline{75.04}   \\
     & \cellcolor{blue!20} $\bigcirc$ & \cellcolor{blue!20} $\bigcirc$ & \cellcolor{blue!20} $\bigcirc$ & \cellcolor{blue!20} $\bigcirc$& \cellcolor{blue!20} $\bigcirc$ & \cellcolor{blue!20} \underline{97.22} & \cellcolor{blue!20} \underline{89.78} & \cellcolor{blue!20} 74.89\\
    \hline
    \multirow{6}{*}{\textbf{VisualBERT}} & \cellcolor{orange!20} $\bigcirc$ & \cellcolor{orange!20} $\times$ & \cellcolor{orange!20} $\times$ & \cellcolor{orange!20} $\times$ & \cellcolor{orange!20} $\times$ & \cellcolor{orange!20} 95.55  & \cellcolor{orange!20} 85.91 & \cellcolor{orange!20} 65.25  \\
    \cline{2-9}
     & $\bigcirc$ & $\times$ & $\bigcirc$ & $\times$ & $\times$ & 95.84 & 85.93& 67.25\\
     & $\bigcirc$ & $\bigcirc$ & $\bigcirc$ & $\times$ & $\times$ & 96.07 & 85.90& 70.14  \\
     & $\bigcirc$ & $\bigcirc$ & $\bigcirc$ & $\times$ & $\bigcirc$ & 96.61 &\underline{87.43} & 70.79 \\
     & $\bigcirc$ & $\bigcirc$ & $\bigcirc$ &$\bigcirc$ &$\times$  & 96.70 &87.00 & 71.28 \\
     & \cellcolor{blue!20} $\bigcirc$ & \cellcolor{blue!20} $\bigcirc$ & \cellcolor{blue!20} $\bigcirc$ & \cellcolor{blue!20} $\bigcirc$ & \cellcolor{blue!20} $\bigcirc$ & \cellcolor{blue!20} \underline{96.73} & \cellcolor{blue!20} 87.18 & \cellcolor{blue!20} \underline{72.65} \\
    \hline
    \multirow{5}{*}{\textbf{LXMERT}} & \cellcolor{orange!20} $\bigcirc$ & \cellcolor{orange!20} $\times$ & \cellcolor{orange!20} $\bigcirc$ & \cellcolor{orange!20} $\times$& \cellcolor{orange!20} $\times$ & \cellcolor{orange!20} \underline{97.83}& \cellcolor{orange!20} 92.29 & \cellcolor{orange!20} 80.51 \\
    \cline{2-9}
     & $\bigcirc$ & $\bigcirc$ & $\bigcirc$ &$\times$ & $\times$ &97.67  & 94.15& 82.80 \\
     & $\bigcirc$ & $\bigcirc$ & $\bigcirc$ &$\bigcirc$ & $\times$ &97.70  & 94.59& 82.60 \\
     & $\bigcirc$ & $\bigcirc$ & $\bigcirc$ & $\times$& $\bigcirc$ &97.74  &94.49 & 83.71\\
     & \cellcolor{blue!20}  $\bigcirc$ & \cellcolor{blue!20} $\bigcirc$ & \cellcolor{blue!20} $\bigcirc$ & \cellcolor{blue!20} $\bigcirc$ & \cellcolor{blue!20} $\bigcirc$ & \cellcolor{blue!20} 97.74 & \cellcolor{blue!20} \underline{95.07} & \cellcolor{blue!20} \underline{84.43}\\
    \hline
    \multirow{4}{*}{\textbf{Our Model}} 
     & \cellcolor{orange!20} $\bigcirc$ & \cellcolor{orange!20} $\bigcirc$ & \cellcolor{orange!20} $\bigcirc$ & \cellcolor{orange!20} $\times$ & \cellcolor{orange!20} $\times$ & \cellcolor{orange!20} 99.06  & \cellcolor{orange!20} 95.07 & \cellcolor{orange!20} 84.56 \\
     \cline{2-9}
     & $\bigcirc$ & $\bigcirc$ & $\bigcirc$ &$\bigcirc$ & $\times$ & \textbf{99.30} & 95.32& 84.75 \\
     & $\bigcirc$ & $\bigcirc$ & $\bigcirc$ & $\times$& $\bigcirc$ &99.12  &95.50 & 85.77\\
     & \cellcolor{blue!20} $\bigcirc$ & \cellcolor{blue!20} $\bigcirc$ & \cellcolor{blue!20} $\bigcirc$ & \cellcolor{blue!20} $\bigcirc$ & 
     \cellcolor{blue!20} $\bigcirc$ & \cellcolor{blue!20} 99.09  & \cellcolor{blue!20} \bf 95.50 & \cellcolor{blue!20} \bf 86.98\\
     \hline
    \end{tabular}
    \end{adjustbox}
    \end{center}
    \caption{Different Background Colours. This is to show the results of \colorbox{orange!20}{vanilla model} (first row of each model group), \colorbox{blue!20}{model with all aspect features} (last row of each model group) on $\mathcal{D}$, $\mathcal{P}$, and $\mathcal{H}$ Test Sets. RoI inputs contain visual ($V$), textual ($T$), positional ($P$), text density ($D$), gap distance ($G$) features, where $\bigcirc$ and $\times$ represent using or not using the specific feature in that column.}
    \label{tab:taskb1_aspects}
    \vspace*{-2em}
\end{table}

M4C initially contains $V$, $T$, and $P$ of input RoIs, and is a non-pretrained model with random initial weights. Thus, it may cause only slight improvements can be found in Table~\ref{tab:taskb1_aspects} after adding $D$ and $G$ into the model on three test sets. For VisualBERT, the vanilla model only contains RoIs visual features; after stacking more aspect features, the performance is gradually improved where the F1-score on $\mathcal{H}$ increases from 65.25\% (vanilla) to \colorbox{blue!20}{72.65\%} (with $V, T, S, D, G$). There is an apparent increase after adding $D$ and $G$ into VisualBERT, which may contribute to the additional features making the input representations more comprehensive. A similar trend also can be found in LXMERTs where $T$ may contribute to a more positive effect (about 2\% increasing) than VisualBERT on $\mathcal{P}$ and $\mathcal{H}$. Regarding our model, although there is no noticeable improvement observed on $\mathcal{D}$, the positive effects on $\mathcal{P}$ and $\mathcal{H}$ can be found, especially for $\mathcal{H}$ (from \colorbox{orange!20}{84.56}\% to \colorbox{blue!20}{86.98\%}). Generally, the proposed additional features could help the model better understand form structures to improve the model generalisation ability, notably when the appearance of input forms differs during the training and inference stage. 

\subsubsection{Positional Encoding Validation}
\begin{table}[t]
    \begin{center}
    \begin{adjustbox}{max width = \linewidth}
    \begin{tabular}{c|c|c|c}
    \hline
    \bf PE Approach & \bf Digital($\mathcal{D}$) & \bf Printed($\mathcal{P}$) & \bf Handwritten($\mathcal{H}$)\\
    \hline
    Without Positional Encoding & 98.83 & 94.34 & 85.98\\
    \hline
    Linear Projection & 99.09 & 95.50 & 86.98\\
    \hline
    \bf XY-Positional Encoding & \bf 99.22 & \bf 96.14 & \bf 89.13 \\
     \hline
    \end{tabular}
    \end{adjustbox}
    \end{center}    \caption{Performance of Various Positional Encodings (PE)}
    \label{tab:pos_emb}
    \vspace*{-2em}
\end{table}
To demonstrate the effectiveness of XY-pos, we conduct external experiments to compare the performance of the model with linear projection PE methods, also without any PE methods. Compared model without any PE in Table~\ref{tab:pos_emb}, linear projection and XY-pos can have an apparent increase in all three test sets. It illustrates that the positional information is significant to understand the layout structure for form understanding. Furthermore, from Table 6, we can find XY-pos reach a better performance than linear projection on $\mathcal{D}$, $\mathcal{P}$ and $\mathcal{H}$. Significantly, the F1 increased from 86.98\% to 89.13\% on $\mathcal{H}$. It demonstrates XY-Pos may capture more positional information to understand various layout structures of input forms better.

\subsubsection{Fine-grained Training Set Ratio}
\begin{figure}[tb]
    \hspace*{0cm}
     \centering
     \hspace*{-0.5em}
     \begin{subfigure}[b]{0.15\textwidth}
         \centering
         \includegraphics[height=2.5cm]{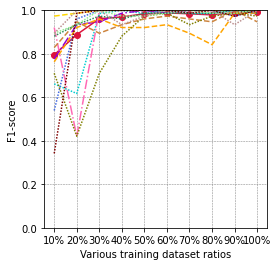}
         \caption{Digital ($\mathcal{D}$)}
         \label{fig:taskb_our_model_digital}
     \end{subfigure}
     \hspace*{-0.8em}
     \begin{subfigure}[b]{0.15\textwidth}
         \centering
         \includegraphics[height=2.5cm]{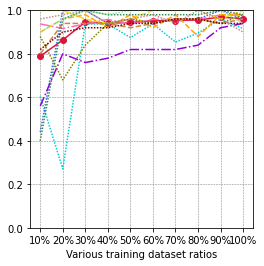}
         \caption{Printed ($\mathcal{P}$)}
         \label{fig:taskb_our_model_printed}
     \end{subfigure}
     \hspace*{-0.5em}
     \begin{subfigure}[b]{0.18\textwidth}
         \centering
         \includegraphics[height=2.5cm]{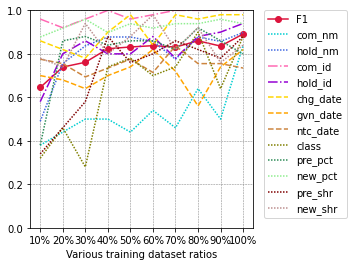}
         \caption{Handwritten ($\mathcal{H}$)}
         \label{fig:taskb_our_model_handwritten}
     \end{subfigure}
        \caption{Performance of our Model with Fine-grained Training Set Ratios on Three Test Sets}
        \label{fig:taskb1_ratio}
         \vspace*{-0.5em}
\end{figure}

For exploring the training size influences for form understanding, we define fine-grained training set ratios (from 10\% to 100\%) to train our model and represent the evaluation performance on $\mathcal{D}$, $\mathcal{P}$ and $\mathcal{H}$ in Figure~\ref{fig:taskb1_ratio}. With increasing training size, fluctuating increases can be observed for overall and breakdown performance on three test sets. The rapid increases can be observed on $\mathcal{D}$ and $\mathcal{P}$ before reaching 50\% $\mathcal{T}$, following the stable trends after reaching this critical ratio. 
However, for handwritten set $\mathcal{H}$, we can find an apparent overall performance increase during the entire training size interval. It may demonstrate the wide variety of our training dataset $\mathcal{H}$ can result in a more generic trained model to extract specific key information more effectively. 
Additionally, unlike the datasets providing general key-value annotations, we can show the training-size sensitivity of specific key-value pairs to analyse the robustness of the trained model with limited training data. For example, \textit{com\_nm}, \textit{com\_id}, and \textit{hold\_nm} show more sensitive to training size, while date-related keys such as \textit{ntc\_date} are less sensitive.

\begin{table}[t]
    \begin{center}
    \begin{adjustbox}{max width = \linewidth}

    \begin{tabular}{c|c|ll|ll|lll|l|ll|ll}
    
    \hline
    \multirow{3}{*}{\textbf{Model}} & \multirow{3}{*}{\textbf{Acc}}  & \multicolumn{12}{c}{\textbf{Breakdowns (Accuracy)}}\\
    \cline{3-14}
    &&  \multicolumn{2}{c|}{\textbf{nm(name)}} & \multicolumn{2}{c|}{\textbf{id}}  &  \multicolumn{3}{c|}{\textbf{date}}  & \multirow{2}{*}{\textbf{class}} & \multicolumn{2}{c|}{\textbf{pct(percent)}}& \multicolumn{2}{c}{\textbf{shr(share)}}\\
    \cline{3-9} \cline{11-14}
    &&  \textbf{com} & \textbf{hold} & \textbf{com} &\textbf{hold} & \textbf{chg} & \textbf{ntc} &\textbf{gvn}  &  &\textbf{pre} & \textbf{new}&\textbf{pre} & \textbf{new}\\
    \hline
    \textbf{LXMERT} & 43.78 & 86.30 & 80.14& 69.18& 52.74 & 72.60 & 65.07 & 54.79& 4.79 & 6.85& 24.66& 5.48 & 2.74\\
    \textbf{VisualBERT} & 57.48 & 88.36 & 78.08& 63.01& 59.59 & 39.04 & 70.55& 37.67 & 47.95 & 50.00& 28.77 & 53.42 & 73.29\\
    \textbf{M4C} & 69.35 & 83.56 & 80.82 & 63.70& 73.29 & 65.75 & 57.53 & 82.88 & 61.64 & 65.75 & 58.90& 59.59& 78.77\\
    \hline
    \textbf{Ours} & 72.72 & 91.78 & 81.51 & 71.92& 71.23 & 80.82 & 80.82 & 84.93 & 71.92 & 70.55 & 80.14& 50.00& 57.53\\
    \hline
    \end{tabular}
    \end{adjustbox}
    \end{center}
    \caption{Overall performance and Breakdown Results of Key Information Extraction Task with PDFminer}
    \label{tab:taskb2_pdfminer}
    \vspace*{-2.5em}
\end{table}

\subsection{Task B with PDF Parser}

\label{sec:case_study1}
Real-world users, especially non-deep learning users, are challenged to get input RoIs through well pretrained layout analysing models because of lacking ground truth annotations and relevant background knowledge. Thus, using textlines extracted by specific PDF parsing tools is an alternative way to obtain the input of adopted well-trained models, such as the green rectangles in Figure~\ref{fig:case_study}. We use PDFminer (a widely used PDF parser \footnote{\url{https://pypi.org/project/pdfminer/}}) to extract RoIs for replacing the ground truth RoIs in a digital set ($\mathcal{D}$) and feed them into three trained baseline models and our models. We use IoU = 0.5 as a threshold to calculate accuracy, as Table~\ref{tab:taskb2_pdfminer} shows. 

Different from the trend of feeding ground truth RoIs during the testing stage, the non-pretrained model M4C can achieve much better accuracy (69.35\%) compared with pre-trained LXMERT (35.24\%) and VisualBERT (53.34\%). The possible reason might be that too many noise RoIs detected by PDFminer feed into the large-scale pretrained models, which decreases those heavy models' RoIs feature representation ability. Our model can achieve the highest overall performance among all tested models. It demonstrates that the proposed XY-pos enhanced dual-level model can improve generality to understand the input contents better. Regarding breakdown results, \textit{com\_nm} and \textit{hold\_nm} could always perform better. Most of the key information extracted by our model could hold higher accuracy than other baselines, such as \textit{com\_id} and \textit{chg\_date}. Notably, LXMERT shows high sensitivity to the bounding box precision of input RoIs. However, because PDFminer extracted bounding boxes are inaccurate, especially in the table area, the keys located in the table are much lower than other models, such as \textit{class}, \textit{pre\_pct}, \textit{pre\_shr}, etc.

\section{Case Study: Transfer Learning} 
\begin{figure}[t]
     \centering
         \includegraphics[width=0.4\textwidth]{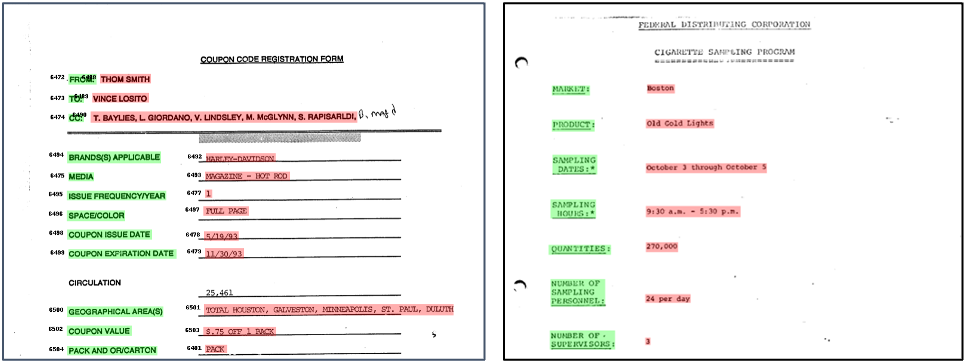}
         \caption{Two Document Samples with Selected \colorbox{green!30}{Question} and \colorbox{red!30}{Answer} Pairs from FUNSD Subset}
         \label{fig:case_study}
         \vspace*{-1.5em}
\end{figure}
In order to check the feasibility of our proposed data and model, we conducted a case study with transfer learning. Note that transfer learning aims to storing knowledge gained while solving our dataset and apply pre-trained models to a different but related problem. In this case study, we applied the best (our model) and second-best models (LXMERT), trained by the Form-NLU dataset, to the publicly available benchmark FUNSD. 
We use the selected key (question)-value (answer) pairs from the FUNSD~\cite{funsd} dataset like the samples shown in Figure~\ref{fig:case_study} \footnote{Please refer to \url{https://github.com/adlnlp/form\_nlu/blob/main/README.md\#case-study-setup} to check the setup detail}. Based on the testing results, even if the nature of FUNSD are entirely different from our Form-NLU dataset, the models trained on our dataset can still achieve sound performance. 
The result shows that LXMERT (the second best of in Task B) correctly predicted 80 (53.33\%) samples out of the 150 FUNSD key(question)-value(answer) pairs while our model achieved \textbf{94 (62.67\%)} correct predictions. It demonstrates that our Form-NLU dataset can learn the general form layout and be applied in other benchmarks, in addition to this, our proposed model is efficient to extract a feature representation for form understanding.

\section{Conclusion}
We proposed Form-NLU, a new form structure understanding and key information extraction dataset. The proposed dataset covers the important point of view, enabling the interpretation of the form designer's specific intention and the alignment of user-written value on it. The dataset includes three form types: digital, printed, and handwritten, which cover diverse form appearances/layouts and deal with their noises. Moreover, we propose a new strong baseline for form structure understanding and key-value information extraction, which applies robust positional and logical relations. Our model outperformed all state-of-the-art models in key-value information extraction tasks. We do hope that our proposed dataset and model can be a great insight into form structure and information analysis, hence, we adopted with off-the-shelf pdf layout extraction tool and also provide its feasibility by conducting transfer learning.

\bibliographystyle{ACM-Reference-Format}
\bibliography{sample-base}


\begin{thebibliography}{32}


\ifx \showCODEN    \undefined \def \showCODEN     #1{\unskip}     \fi
\ifx \showDOI      \undefined \def \showDOI       #1{#1}\fi
\ifx \showISBNx    \undefined \def \showISBNx     #1{\unskip}     \fi
\ifx \showISBNxiii \undefined \def \showISBNxiii  #1{\unskip}     \fi
\ifx \showISSN     \undefined \def \showISSN      #1{\unskip}     \fi
\ifx \showLCCN     \undefined \def \showLCCN      #1{\unskip}     \fi
\ifx \shownote     \undefined \def \shownote      #1{#1}          \fi
\ifx \showarticletitle \undefined \def \showarticletitle #1{#1}   \fi
\ifx \showURL      \undefined \def \showURL       {\relax}        \fi
\providecommand\bibfield[2]{#2}
\providecommand\bibinfo[2]{#2}
\providecommand\natexlab[1]{#1}
\providecommand\showeprint[2][]{arXiv:#2}

\bibitem[Chang et~al\mbox{.}(2016)]%
        {chang2016informativeness}
\bibfield{author}{\bibinfo{person}{Millicent Chang}, \bibinfo{person}{Raymond
  da Silva~Rosa}, {and} \bibinfo{person}{Wilson Ng}.}
  \bibinfo{year}{2016}\natexlab{}.
\newblock \showarticletitle{The Informativeness of Substantial Shareholder
  Trading in the Lead up to a Takeover Bid}. In \bibinfo{booktitle}{\emph{Asian
  Finance Association (AsianFA) 2016 Conference}}.
\newblock


\bibitem[Chi et~al\mbox{.}(2019)]%
        {scitsr}
\bibfield{author}{\bibinfo{person}{Zewen Chi}, \bibinfo{person}{Heyan Huang},
  \bibinfo{person}{Heng-Da Xu}, \bibinfo{person}{Houjin Yu},
  \bibinfo{person}{Wanxuan Yin}, {and} \bibinfo{person}{Xian-Ling Mao}.}
  \bibinfo{year}{2019}\natexlab{}.
\newblock \showarticletitle{Complicated table structure recognition}.
\newblock \bibinfo{journal}{\emph{arXiv preprint arXiv:1908.04729}}
  (\bibinfo{year}{2019}).
\newblock


\bibitem[Cohen(1960)]%
        {cohen1960coefficient}
\bibfield{author}{\bibinfo{person}{Jacob Cohen}.}
  \bibinfo{year}{1960}\natexlab{}.
\newblock \showarticletitle{A coefficient of agreement for nominal scales}.
\newblock \bibinfo{journal}{\emph{Educational and psychological measurement}}
  \bibinfo{volume}{20}, \bibinfo{number}{1} (\bibinfo{year}{1960}),
  \bibinfo{pages}{37--46}.
\newblock


\bibitem[Ding et~al\mbox{.}(2022)]%
        {vdoc}
\bibfield{author}{\bibinfo{person}{Yihao Ding}, \bibinfo{person}{Zhe Huang},
  \bibinfo{person}{Runlin Wang}, \bibinfo{person}{YanHang Zhang},
  \bibinfo{person}{Xianru Chen}, \bibinfo{person}{Yuzhong Ma},
  \bibinfo{person}{Hyunsuk Chung}, {and} \bibinfo{person}{Soyeon~Caren Han}.}
  \bibinfo{year}{2022}\natexlab{}.
\newblock \showarticletitle{V-Doc: Visual questions answers with Documents}. In
  \bibinfo{booktitle}{\emph{Proceedings of the IEEE/CVF Conference on Computer
  Vision and Pattern Recognition}}. \bibinfo{pages}{21492--21498}.
\newblock


\bibitem[Gu et~al\mbox{.}(2022)]%
        {gu2022xylayoutlm}
\bibfield{author}{\bibinfo{person}{Zhangxuan Gu}, \bibinfo{person}{Changhua
  Meng}, \bibinfo{person}{Ke Wang}, \bibinfo{person}{Jun Lan},
  \bibinfo{person}{Weiqiang Wang}, \bibinfo{person}{Ming Gu}, {and}
  \bibinfo{person}{Liqing Zhang}.} \bibinfo{year}{2022}\natexlab{}.
\newblock \showarticletitle{Xylayoutlm: Towards layout-aware multimodal
  networks for visually-rich document understanding}. In
  \bibinfo{booktitle}{\emph{Proceedings of the IEEE/CVF Conference on Computer
  Vision and Pattern Recognition}}. \bibinfo{pages}{4583--4592}.
\newblock


\bibitem[He et~al\mbox{.}(2017)]%
        {maskrcnn}
\bibfield{author}{\bibinfo{person}{Kaiming He}, \bibinfo{person}{Georgia
  Gkioxari}, \bibinfo{person}{Piotr Doll{\'a}r}, {and} \bibinfo{person}{Ross
  Girshick}.} \bibinfo{year}{2017}\natexlab{}.
\newblock \showarticletitle{Mask r-cnn}. In
  \bibinfo{booktitle}{\emph{Proceedings of the IEEE international conference on
  computer vision}}. \bibinfo{pages}{2961--2969}.
\newblock


\bibitem[Hu et~al\mbox{.}(2020)]%
        {m4c}
\bibfield{author}{\bibinfo{person}{Ronghang Hu}, \bibinfo{person}{Amanpreet
  Singh}, \bibinfo{person}{Trevor Darrell}, {and} \bibinfo{person}{Marcus
  Rohrbach}.} \bibinfo{year}{2020}\natexlab{}.
\newblock \showarticletitle{Iterative answer prediction with pointer-augmented
  multimodal transformers for textvqa}. In
  \bibinfo{booktitle}{\emph{Proceedings of the IEEE/CVF Conference on Computer
  Vision and Pattern Recognition}}. \bibinfo{pages}{9992--10002}.
\newblock


\bibitem[Huang et~al\mbox{.}(2022)]%
        {huang2022layoutlmv3}
\bibfield{author}{\bibinfo{person}{Yupan Huang}, \bibinfo{person}{Tengchao Lv},
  \bibinfo{person}{Lei Cui}, \bibinfo{person}{Yutong Lu}, {and}
  \bibinfo{person}{Furu Wei}.} \bibinfo{year}{2022}\natexlab{}.
\newblock \showarticletitle{LayoutLMv3: Pre-training for Document AI with
  Unified Text and Image Masking}.
\newblock \bibinfo{journal}{\emph{arXiv preprint arXiv:2204.08387}}
  (\bibinfo{year}{2022}).
\newblock


\bibitem[Huang et~al\mbox{.}(2019)]%
        {scoie}
\bibfield{author}{\bibinfo{person}{Zheng Huang}, \bibinfo{person}{Kai Chen},
  \bibinfo{person}{Jianhua He}, \bibinfo{person}{Xiang Bai},
  \bibinfo{person}{Dimosthenis Karatzas}, \bibinfo{person}{Shijian Lu}, {and}
  \bibinfo{person}{CV Jawahar}.} \bibinfo{year}{2019}\natexlab{}.
\newblock \showarticletitle{Icdar 2019 robust reading challenge on scanned
  receipts ocr and information extraction}. In
  \bibinfo{booktitle}{\emph{International Conference on Document Analysis
  Recognition}}.
\newblock


\bibitem[Jaume et~al\mbox{.}(2019)]%
        {funsd}
\bibfield{author}{\bibinfo{person}{Guillaume Jaume},
  \bibinfo{person}{Hazim~Kemal Ekenel}, {and} \bibinfo{person}{Jean-Philippe
  Thiran}.} \bibinfo{year}{2019}\natexlab{}.
\newblock \showarticletitle{Funsd: A dataset for form understanding in noisy
  scanned documents}. In \bibinfo{booktitle}{\emph{2019 International
  Conference on Document Analysis and Recognition Workshops (ICDARW)}},
  Vol.~\bibinfo{volume}{2}. IEEE, \bibinfo{pages}{1--6}.
\newblock


\bibitem[Kim et~al\mbox{.}(2021)]%
        {kim2021vilt}
\bibfield{author}{\bibinfo{person}{Wonjae Kim}, \bibinfo{person}{Bokyung Son},
  {and} \bibinfo{person}{Ildoo Kim}.} \bibinfo{year}{2021}\natexlab{}.
\newblock \showarticletitle{Vilt: Vision-and-language transformer without
  convolution or region supervision}. In
  \bibinfo{booktitle}{\emph{International Conference on Machine Learning}}.
  PMLR, \bibinfo{pages}{5583--5594}.
\newblock


\bibitem[Li et~al\mbox{.}(2019)]%
        {visualbert}
\bibfield{author}{\bibinfo{person}{Liunian~Harold Li}, \bibinfo{person}{Mark
  Yatskar}, \bibinfo{person}{Da Yin}, \bibinfo{person}{Cho-Jui Hsieh}, {and}
  \bibinfo{person}{Kai-Wei Chang}.} \bibinfo{year}{2019}\natexlab{}.
\newblock \showarticletitle{Visualbert: A simple and performant baseline for
  vision and language}.
\newblock \bibinfo{journal}{\emph{arXiv preprint arXiv:1908.03557}}
  (\bibinfo{year}{2019}).
\newblock


\bibitem[Li et~al\mbox{.}(2020)]%
        {docbank}
\bibfield{author}{\bibinfo{person}{Minghao Li}, \bibinfo{person}{Yiheng Xu},
  \bibinfo{person}{Lei Cui}, \bibinfo{person}{Shaohan Huang},
  \bibinfo{person}{Furu Wei}, \bibinfo{person}{Zhoujun Li}, {and}
  \bibinfo{person}{Ming Zhou}.} \bibinfo{year}{2020}\natexlab{}.
\newblock \showarticletitle{DocBank: A Benchmark Dataset for Document Layout
  Analysis}. In \bibinfo{booktitle}{\emph{Proceedings of the 28th International
  Conference on Computational Linguistics}}. \bibinfo{pages}{949--960}.
\newblock


\bibitem[Luo et~al\mbox{.}(2022)]%
        {docgcn}
\bibfield{author}{\bibinfo{person}{Siwen Luo}, \bibinfo{person}{Yihao Ding},
  \bibinfo{person}{Siqu Long}, \bibinfo{person}{Josiah Poon}, {and}
  \bibinfo{person}{Soyeon~Caren Han}.} \bibinfo{year}{2022}\natexlab{}.
\newblock \showarticletitle{Doc-GCN: Heterogeneous Graph Convolutional Networks
  for Document Layout Analysis}. In \bibinfo{booktitle}{\emph{Proceedings of
  the 29th International Conference on Computational Linguistics}}.
  \bibinfo{pages}{2906--2916}.
\newblock


\bibitem[Mathew et~al\mbox{.}(2021)]%
        {docvqa}
\bibfield{author}{\bibinfo{person}{Minesh Mathew}, \bibinfo{person}{Dimosthenis
  Karatzas}, {and} \bibinfo{person}{CV Jawahar}.}
  \bibinfo{year}{2021}\natexlab{}.
\newblock \showarticletitle{Docvqa: A dataset for vqa on document images}. In
  \bibinfo{booktitle}{\emph{Proceedings of the IEEE/CVF Winter Conference on
  Applications of Computer Vision}}. \bibinfo{pages}{2200--2209}.
\newblock


\bibitem[Park et~al\mbox{.}(2019)]%
        {cord}
\bibfield{author}{\bibinfo{person}{Seunghyun Park}, \bibinfo{person}{Seung
  Shin}, \bibinfo{person}{Bado Lee}, \bibinfo{person}{Junyeop Lee},
  \bibinfo{person}{Jaeheung Surh}, \bibinfo{person}{Minjoon Seo}, {and}
  \bibinfo{person}{Hwalsuk Lee}.} \bibinfo{year}{2019}\natexlab{}.
\newblock \showarticletitle{CORD: a consolidated receipt dataset for post-OCR
  parsing}. In \bibinfo{booktitle}{\emph{Workshop on Document Intelligence at
  NeurIPS 2019}}.
\newblock


\bibitem[Rausch et~al\mbox{.}(2021)]%
        {docparser}
\bibfield{author}{\bibinfo{person}{Johannes Rausch}, \bibinfo{person}{Octavio
  Martinez}, \bibinfo{person}{Fabian Bissig}, \bibinfo{person}{Ce Zhang}, {and}
  \bibinfo{person}{Stefan Feuerriegel}.} \bibinfo{year}{2021}\natexlab{}.
\newblock \showarticletitle{Docparser: Hierarchical document structure parsing
  from renderings}. In \bibinfo{booktitle}{\emph{Proceedings of the AAAI
  Conference on Artificial Intelligence}}, Vol.~\bibinfo{volume}{35}.
  \bibinfo{pages}{4328--4338}.
\newblock


\bibitem[Ren et~al\mbox{.}(2015)]%
        {fasterrcnn}
\bibfield{author}{\bibinfo{person}{Shaoqing Ren}, \bibinfo{person}{Kaiming He},
  \bibinfo{person}{Ross Girshick}, {and} \bibinfo{person}{Jian Sun}.}
  \bibinfo{year}{2015}\natexlab{}.
\newblock \showarticletitle{Faster r-cnn: Towards real-time object detection
  with region proposal networks}.
\newblock \bibinfo{journal}{\emph{Advances in neural information processing
  systems}}  \bibinfo{volume}{28} (\bibinfo{year}{2015}).
\newblock


\bibitem[Stanis{\l}awek et~al\mbox{.}(2021)]%
        {kleister}
\bibfield{author}{\bibinfo{person}{Tomasz Stanis{\l}awek},
  \bibinfo{person}{Filip Grali{\'n}ski}, \bibinfo{person}{Anna Wr{\'o}blewska},
  \bibinfo{person}{Dawid Lipi{\'n}ski}, \bibinfo{person}{Agnieszka Kaliska},
  \bibinfo{person}{Paulina Rosalska}, \bibinfo{person}{Bartosz Topolski}, {and}
  \bibinfo{person}{Przemys{\l}aw Biecek}.} \bibinfo{year}{2021}\natexlab{}.
\newblock \showarticletitle{Kleister: key information extraction datasets
  involving long documents with complex layouts}. In
  \bibinfo{booktitle}{\emph{International Conference on Document Analysis and
  Recognition}}. Springer, \bibinfo{pages}{564--579}.
\newblock


\bibitem[Tan and Bansal(2019)]%
        {tan2019lxmert}
\bibfield{author}{\bibinfo{person}{Hao Tan} {and} \bibinfo{person}{Mohit
  Bansal}.} \bibinfo{year}{2019}\natexlab{}.
\newblock \showarticletitle{Lxmert: Learning cross-modality encoder
  representations from transformers}.
\newblock \bibinfo{journal}{\emph{arXiv preprint arXiv:1908.07490}}
  (\bibinfo{year}{2019}).
\newblock


\bibitem[Tanaka et~al\mbox{.}(2021)]%
        {visualmrc}
\bibfield{author}{\bibinfo{person}{Ryota Tanaka}, \bibinfo{person}{Kyosuke
  Nishida}, {and} \bibinfo{person}{Sen Yoshida}.}
  \bibinfo{year}{2021}\natexlab{}.
\newblock \showarticletitle{VisualMRC: Machine Reading Comprehension on
  Document Images}. In \bibinfo{booktitle}{\emph{Proceedings of the AAAI
  Conference on Artificial Intelligence}}, Vol.~\bibinfo{volume}{35}.
  \bibinfo{pages}{13878--13888}.
\newblock


\bibitem[Vaswani et~al\mbox{.}(2017)]%
        {transformer}
\bibfield{author}{\bibinfo{person}{Ashish Vaswani}, \bibinfo{person}{Noam
  Shazeer}, \bibinfo{person}{Niki Parmar}, \bibinfo{person}{Jakob Uszkoreit},
  \bibinfo{person}{Llion Jones}, \bibinfo{person}{Aidan~N Gomez},
  \bibinfo{person}{{\L}ukasz Kaiser}, {and} \bibinfo{person}{Illia
  Polosukhin}.} \bibinfo{year}{2017}\natexlab{}.
\newblock \showarticletitle{Attention is all you need}.
\newblock \bibinfo{journal}{\emph{Advances in neural information processing
  systems}}  \bibinfo{volume}{30} (\bibinfo{year}{2017}).
\newblock


\bibitem[Vinyals et~al\mbox{.}(2015)]%
        {vinyals2015pointer}
\bibfield{author}{\bibinfo{person}{Oriol Vinyals}, \bibinfo{person}{Meire
  Fortunato}, {and} \bibinfo{person}{Navdeep Jaitly}.}
  \bibinfo{year}{2015}\natexlab{}.
\newblock \showarticletitle{Pointer networks}.
\newblock \bibinfo{journal}{\emph{Advances in neural information processing
  systems}}  \bibinfo{volume}{28} (\bibinfo{year}{2015}).
\newblock


\bibitem[Wang et~al\mbox{.}(2021)]%
        {ephoie}
\bibfield{author}{\bibinfo{person}{Jiapeng Wang}, \bibinfo{person}{Chongyu
  Liu}, \bibinfo{person}{Lianwen Jin}, \bibinfo{person}{Guozhi Tang},
  \bibinfo{person}{Jiaxin Zhang}, \bibinfo{person}{Shuaitao Zhang},
  \bibinfo{person}{Qianying Wang}, \bibinfo{person}{Yaqiang Wu}, {and}
  \bibinfo{person}{Mingxiang Cai}.} \bibinfo{year}{2021}\natexlab{}.
\newblock \showarticletitle{Towards robust visual information extraction in
  real world: new dataset and novel solution}. In
  \bibinfo{booktitle}{\emph{Proceedings of the AAAI Conference on Artificial
  Intelligence}}, Vol.~\bibinfo{volume}{35}. \bibinfo{pages}{2738--2745}.
\newblock


\bibitem[Xu et~al\mbox{.}(2020)]%
        {xu2020layoutlm}
\bibfield{author}{\bibinfo{person}{Yiheng Xu}, \bibinfo{person}{Minghao Li},
  \bibinfo{person}{Lei Cui}, \bibinfo{person}{Shaohan Huang},
  \bibinfo{person}{Furu Wei}, {and} \bibinfo{person}{Ming Zhou}.}
  \bibinfo{year}{2020}\natexlab{}.
\newblock \showarticletitle{Layoutlm: Pre-training of text and layout for
  document image understanding}. In \bibinfo{booktitle}{\emph{Proceedings of
  the 26th ACM SIGKDD International Conference on Knowledge Discovery \& Data
  Mining}}. \bibinfo{pages}{1192--1200}.
\newblock


\bibitem[Xu et~al\mbox{.}(2021a)]%
        {xu2021layoutxlm}
\bibfield{author}{\bibinfo{person}{Yiheng Xu}, \bibinfo{person}{Tengchao Lv},
  \bibinfo{person}{Lei Cui}, \bibinfo{person}{Guoxin Wang},
  \bibinfo{person}{Yijuan Lu}, \bibinfo{person}{Dinei Florencio},
  \bibinfo{person}{Cha Zhang}, {and} \bibinfo{person}{Furu Wei}.}
  \bibinfo{year}{2021}\natexlab{a}.
\newblock \showarticletitle{Layoutxlm: Multimodal pre-training for multilingual
  visually-rich document understanding}.
\newblock \bibinfo{journal}{\emph{arXiv preprint arXiv:2104.08836}}
  (\bibinfo{year}{2021}).
\newblock


\bibitem[Xu et~al\mbox{.}(2022)]%
        {xfund}
\bibfield{author}{\bibinfo{person}{Yiheng Xu}, \bibinfo{person}{Tengchao Lv},
  \bibinfo{person}{Lei Cui}, \bibinfo{person}{Guoxin Wang},
  \bibinfo{person}{Yijuan Lu}, \bibinfo{person}{Dinei Florencio},
  \bibinfo{person}{Cha Zhang}, {and} \bibinfo{person}{Furu Wei}.}
  \bibinfo{year}{2022}\natexlab{}.
\newblock \showarticletitle{XFUND: A Benchmark Dataset for Multilingual
  Visually Rich Form Understanding}. In \bibinfo{booktitle}{\emph{Findings of
  the Association for Computational Linguistics: ACL 2022}}.
  \bibinfo{pages}{3214--3224}.
\newblock


\bibitem[Xu et~al\mbox{.}(2021b)]%
        {xu2021layoutlmv2}
\bibfield{author}{\bibinfo{person}{Yang Xu}, \bibinfo{person}{Yiheng Xu},
  \bibinfo{person}{Tengchao Lv}, \bibinfo{person}{Lei Cui},
  \bibinfo{person}{Furu Wei}, \bibinfo{person}{Guoxin Wang},
  \bibinfo{person}{Yijuan Lu}, \bibinfo{person}{Dinei Florencio},
  \bibinfo{person}{Cha Zhang}, \bibinfo{person}{Wanxiang Che}, {et~al\mbox{.}}}
  \bibinfo{year}{2021}\natexlab{b}.
\newblock \showarticletitle{LayoutLMv2: Multi-modal Pre-training for
  Visually-rich Document Understanding}. In
  \bibinfo{booktitle}{\emph{Proceedings of the 59th Annual Meeting of the
  Association for Computational Linguistics and the 11th International Joint
  Conference on Natural Language Processing (Volume 1: Long Papers)}}.
  \bibinfo{pages}{2579--2591}.
\newblock


\bibitem[Zhang et~al\mbox{.}(2021)]%
        {zhang2021entity}
\bibfield{author}{\bibinfo{person}{Yue Zhang}, \bibinfo{person}{Bo Zhang},
  \bibinfo{person}{Rui Wang}, \bibinfo{person}{Junjie Cao},
  \bibinfo{person}{Chen Li}, {and} \bibinfo{person}{Zuyi Bao}.}
  \bibinfo{year}{2021}\natexlab{}.
\newblock \showarticletitle{Entity Relation Extraction as Dependency Parsing in
  Visually Rich Documents}.
\newblock \bibinfo{journal}{\emph{arXiv preprint arXiv:2110.09915}}
  (\bibinfo{year}{2021}).
\newblock


\bibitem[Zheng et~al\mbox{.}(2021)]%
        {fintabnet}
\bibfield{author}{\bibinfo{person}{Xinyi Zheng}, \bibinfo{person}{Douglas
  Burdick}, \bibinfo{person}{Lucian Popa}, \bibinfo{person}{Xu Zhong}, {and}
  \bibinfo{person}{Nancy Xin~Ru Wang}.} \bibinfo{year}{2021}\natexlab{}.
\newblock \showarticletitle{Global table extractor (gte): A framework for joint
  table identification and cell structure recognition using visual context}. In
  \bibinfo{booktitle}{\emph{Proceedings of the IEEE/CVF winter conference on
  applications of computer vision}}. \bibinfo{pages}{697--706}.
\newblock


\bibitem[Zhong et~al\mbox{.}(2020)]%
        {pubtabnet}
\bibfield{author}{\bibinfo{person}{Xu Zhong}, \bibinfo{person}{Elaheh
  ShafieiBavani}, {and} \bibinfo{person}{Antonio Jimeno~Yepes}.}
  \bibinfo{year}{2020}\natexlab{}.
\newblock \showarticletitle{Image-based table recognition: data, model, and
  evaluation}. In \bibinfo{booktitle}{\emph{European Conference on Computer
  Vision}}. Springer, \bibinfo{pages}{564--580}.
\newblock


\bibitem[Zhong et~al\mbox{.}(2019)]%
        {publaynet}
\bibfield{author}{\bibinfo{person}{Xu Zhong}, \bibinfo{person}{Jianbin Tang},
  {and} \bibinfo{person}{Antonio~Jimeno Yepes}.}
  \bibinfo{year}{2019}\natexlab{}.
\newblock \showarticletitle{Publaynet: largest dataset ever for document layout
  analysis}. In \bibinfo{booktitle}{\emph{2019 International Conference on
  Document Analysis and Recognition (ICDAR)}}. IEEE,
  \bibinfo{pages}{1015--1022}.
\newblock


\end{thebibliography}


\end{document}